\documentclass[sigplan,screen]{acmart} %,review,anonymous
\AtBeginDocument{%
  }

\setcopyright{acmlicensed}
\copyrightyear{2025}
\acmYear{2025}
\acmDOI{XXXXXXX.XXXXXXX}
\acmConference[AAAI'26]{}{Jan.20-27}{Singapore}
\acmISBN{978-1-4503-XXXX-X/2025/09}

\usepackage{graphicx}
\usepackage{float}
\usepackage{subfigure}
\usepackage{enumitem}
\usepackage{multirow}
\begin{document}
\renewcommand{\thesubfigure}{\textbf{(\alph{subfigure})}}  % 使整个（a）加粗
\renewcommand{\thesubtable}{\small{\textbf{(\alph{subtable})}}}  % 使整个（a）加粗
\title{Understanding the Information Cocoon: A Multidimensional Assessment and Analysis of News Recommendation Systems}
\author{Xin Wang}
\affiliation{%
  \institution{School of Computer Science and Technology, Beijing Jiaotong University}
  \city{Beijing}
  \country{China}
}
\email{24125278@bjtu.edu.cn}

\author{Xiaowen Huang}
\authornote{Corresponding author}
\affiliation{%
  \institution{School of Computer Science and Technology, Beijing Jiaotong University}{Beijing Key Laboratory of Traffic Data Mining and Embodied Intelligence}{Key Laboratory of Big Data \& Artificial Intelligence in Transportation, Ministry of Education}
  \city{Beijing}
  \country{China}
}
\email{xwhuang@bjtu.edu.cn}

\author{Jitao Sang}
\affiliation{%
  \institution{School of Computer Science and Technology, Beijing Jiaotong University}{Beijing Key Laboratory of Traffic Data Mining and Embodied Intelligence}{Key Laboratory of Big Data \& Artificial Intelligence in Transportation, Ministry of Education}
  \city{Beijing}
  \country{China}
}
\email{jtsang@bjtu.edu.cn}

\renewcommand{\shortauthors}{Xin Wang, Xiaowen Huang, and Jitao Sang}

\begin{abstract}
Personalized news recommendation systems inadvertently create information cocoons—homogeneous information bubbles that reinforce user biases and amplify societal polarization. To address the lack of comprehensive assessment frameworks in prior research, we propose a multidimensional analysis that evaluates cocoons through dual perspectives: (1) Individual homogenization via topic diversity (including the number of topic categories and category information entropy) and click repetition; (2) Group polarization via network density and community openness. Through multi-round experiments on real-world datasets, we benchmark seven algorithms and reveal critical insights. Furthermore, we design five lightweight mitigation strategies. This work establishes the first unified metric framework for information cocoons and delivers deployable solutions for ethical recommendation systems.
\end{abstract}




\maketitle

% 情感极化分析图
\begin{figure}[t]
 \centering
 \includegraphics[width=1\linewidth]
 {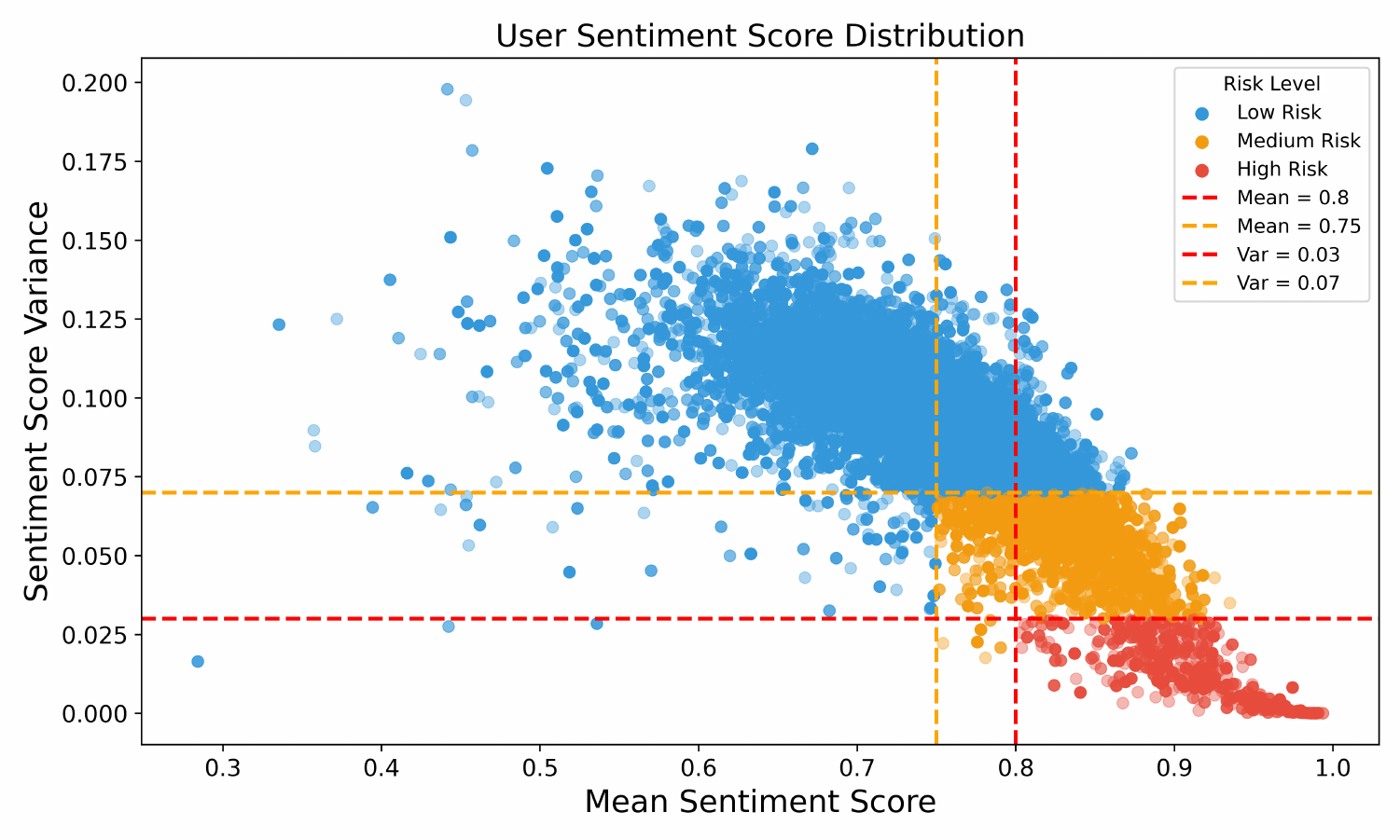}
 \caption{Sentiment analysis of users' clicked news. The higher means and the lower variance imply greater emotional polarization risk.}
 \label{Fig.sentiment_risk}
\end{figure}

\section{Introduction}
With the rapid growth of the digital era, intelligent recommendation technology has matured and is widely applied in various information services, such as social media \cite{mckay2022turn,Avin2024}, news \cite{zhang2023evolution}, short video platforms \cite{Li_2022}, and music platforms \cite{Zeng_2024musicrec}, providing more accurate and personalized information. Among them, news recommendation systems, as authoritative channels of information dissemination, have a real-time and profound impact on public opinion and social perception. As recommendation algorithms continue to improve and personalization deepens, the effect of the information cocoon has gradually emerged. 
%socialmedia-pnas2021
%news Dwivedi2016ASO

The information cocoon \cite{sunstein2001republic} refers to the process in which users are exposed only to content that aligns with their own views \cite{Kalimeris-kdd2021}. Through long-term learning, news recommendation systems tend to reinforce users’ existing preferences, gradually trapping them in a cocoon of homogeneous information and limiting their exposure to diverse viewpoints. This can lead to cognitive bias, emotional polarization, and informational silos \cite{bellina2023effect}. Further, we conducted a statistical analysis of sentiment scores associated with users' historical news clicks and assessed their risk of emotional polarization, as shown in Figure \ref{Fig.sentiment_risk}. We found that a substantial portion of users repeatedly clicked on emotionally charged news articles with similar scores indicating a high risk of polarization. These findings highlight the importance of understanding the formation mechanisms and influencing factors of information cocoons in news recommendation systems, which is essential for fostering an open, diverse and stable digital information environment.

Existing research primarily focuses on detecting and mitigating information cocoons \cite{Anwar-2024}. Some works explore the emergence and development mechanisms of information cocoons \cite{Piao2023,sukiennik2024uncovering}. Several works have attempted to study the information cocoon effect at the individual user level \cite{EDUA-2021,SSLE-individual-2022,Michiels-2023,Li-2022short_video,gu2024modeling}. %IDSR-2020
They proposed some diversified methods to mitigate the cocoon effect \cite{zheng2024facetcrs}. From the group perspective, several studies designed community recommendation algorithms that were aware of the echo chamber \cite{Antonela-2021,Tim-2021,cinus2022effect,liu2024formal}. %Zheng-2021 Grossetti-2019,
Some works also promote diversity through causal reasoning and reinforcement learning methods to alleviate information cocoons \cite{UCRS-2022,stamenkovic2022choosing,CIRS-2023}. 
%From a practical perspective, \citeauthor{CIRS-2023}\shortcite{CIRS-2023} proposed a counterfactual interactive recommendation system that used reinforcement learning and causal reasoning to model user satisfaction, increasing the information diversity \cite{UCRS-2022}. 
However, current studies typically assess the information cocoon effect based on only one or two indicators from a single perspective, leading to a narrow and partial understanding of the issue \cite{feature-aware-KDD2022,Zhang_2024,WANG2024101216}. There is a lack of a comprehensive assessment framework for information cocoons, and the metrics used have not undergone extensive empirical testing, resulting in unreliable mitigation strategies. Moreover, few studies have systematically compared the performance of different models in terms of information cocoons or examined mitigation strategies. Most existing studies are based on single-round recommendations, leaving the long-term evolution of cocoons largely unexplored.

This study investigates the effect of information cocoons within the context of news recommendation systems. By using real-world news datasets with different scales, the research provides a comprehensive analysis of the information cocoon effect, examining its manifestations from both individual and social group perspectives \cite{hu2022ai}. 
At the individual level, the study focuses on the narrowing of users' information scope and the increasing polarization of their opinions, which are quantified through diversity metrics applied to the recommendation list. 
At the group level, the focus shifts to network polarization and group opinion polarization, utilizing indicators related to user-item networks. 
Based on these metrics, we conduct a multidimensional assessment and analysis of the information cocoon effect in multiple classic news recommendation models throughout the recommendation process, and design some lightweight strategies to try to alleviate information cocoons. % The study seeks to explore the formation of information cocoons and the impact of different algorithms on their intensity. 
Through experimentation, the study compares the degree of information cocooning across various recommendation algorithms from multiple perspectives, providing a scientific foundation for the design and improvement of news recommendation models, ultimately contributing to the mitigation of information cocoons and the improvement of information diversity.

The main contribution of our paper is as follows.
\begin{itemize} [topsep=0pt, left=0pt]
\item \textbf{Comprehensive Analysis Framework}: This study develops a comprehensive assessment framework to analyze the information cocoon effect in news recommendation systems. Unlike previous research that relies on limited metrics, this study examines the effect from both individual and group perspectives, providing a more comprehensive understanding of the issue.
\item \textbf{Multidimensional Assessment of Information Cocoons}: The research conducts a multidimensional assessment of the information cocoon effect using real-world news datasets of different scales. It quantitatively evaluates the narrowing of user information scope and opinion polarization at the individual level, as well as network polarization and group opinion polarization at the group level. This approach offers a more detailed and reliable assessment of the information cocoon effect.
% \item \textbf{Investigation of Formation Mechanisms}: This study investigates the formation mechanisms of information cocoons in news recommendation systems. By examining the impact of different algorithmic mechanisms on the intensity of cocoons, the research provides insights into how recommendation algorithms contribute to the entrenchment of homogeneous information bubbles.
\item \textbf{Empirical Comparison of Recommendation Algorithms}: The study compares the degree of information cocooning across various recommendation algorithms through rigorous experimentation, offering insight into their role in reinforcing homogeneous bubbles. The findings provide a robust scientific foundation for the design and optimization of news recommendation models, ultimately contributing to the mitigation of information cocoons and the improvement of information diversity. 
\item \textbf{Evaluation of Mitigation Strategies}: The study designs several lightweight mitigation strategies from both individual and group perspectives, including controlled randomness, model-level regularization and post-hoc re-ranking. These strategies are empirically evaluated using multiple metrics, offering practical guidance for the development of future news recommendation systems aimed at promoting information diversity. 
\end{itemize}

\section{Definition of Information Cocoon}
Regarding the effect of the information cocoon, the related research also contains two concepts with similar meanings, namely the Filter Bubble and Echo Chamber.

\subsection{Filter Bubble}
Filter bubble in recommender systems was proposed by \citeauthor{Nguyen_2014}\shortcite{Nguyen_2014}. It refers to the selective limitation of information access caused by algorithmic filtering mechanisms, such that the user is primarily exposed to content that aligns with their pre-existing beliefs, preferences, and behaviors. This phenomenon is predominantly fueled by personalized recommendation systems, wherein filtering algorithms prioritize relevance (i.e., content aligned with users’ historical interactions) at the expense of diversity. As a result, the system increasingly recommends homogeneous content, leading to a narrowing of the user's information exposure.

Consequently, the filter bubble restricts the diversity of information that individuals encounter, reducing their chances of being exposed to alternative ideas, critical perspectives, or new information that could challenge their existing understanding. This selective exposure can exacerbate group polarization and increase user biases. This stresses the potential negative consequences of personalized recommendation systems, particularly the homogenization of individual recommendations.

\subsection{Echo Chamber}
Echo chamber refers to a social group phenomenon in which people are exposed to information, opinions, and content that reinforce their existing beliefs or biases, within a specific group or community \cite{Nguyen_2020}. Unlike the filter bubble, which operates at the individual level, echo chamber is primarily characterized by the formation of homogeneous groups or communities that share a common stance or viewpoint. These groups typically increase their own beliefs and opinions while minimizing or disregarding opposing perspectives. As a result, echo chamber members are often isolated from diverse viewpoints \cite{Bakshy_2015}, which can lead to the entrenchment of their views and greater group polarization.

In the recommendation system, the algorithms and social networks tend to concentrate individuals with similar opinions in like-minded communities. This leads to the amplification of particular ideologies or perspectives and can contribute to the reinforcement of social divisions.

% 架构图
\begin{figure}[h]
 \centering
 \includegraphics[width=1\linewidth]{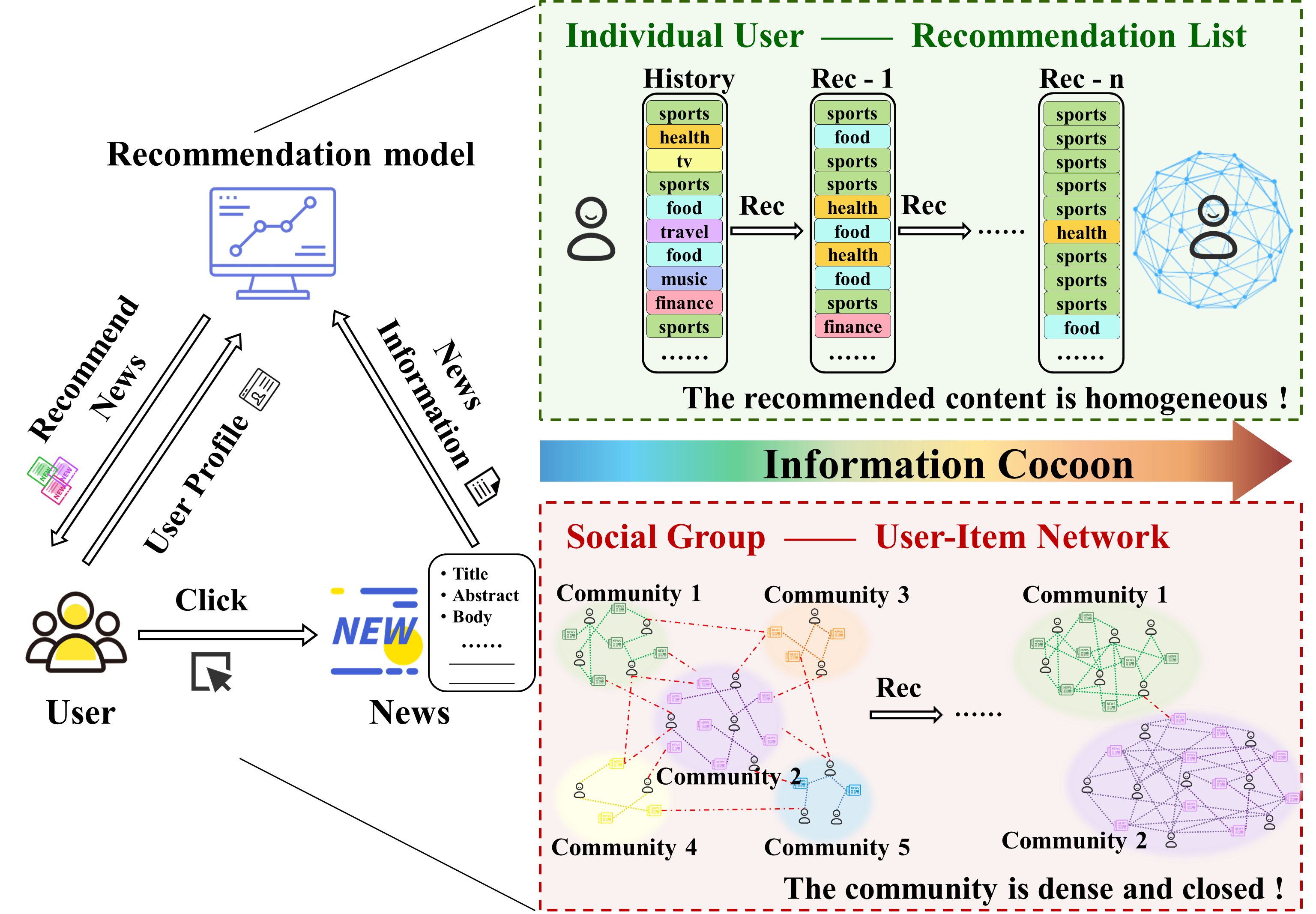}
 \caption{Information cocoon effect in news recommendations. From an individual perspective, it appears as reduced diversity in recommendation lists. From a group perspective, it manifests as network clustering and closure.}
 \label{Fig.architecture}
\end{figure}

\subsection{Information Cocoon}
Information cocoon was first introduced by Sunstein  \cite{sunstein2001republic}. He proposed that, with the advancement of internet technology and the explosion of information, people can freely choose the topics they wish to follow. When individuals are confined within this self-constructed information environment, their lives inevitably become more routine and formulaic. Over time, they will be trapped in an isolated ``cocoon" of their own making.

It can be observed that, compared to the previous two concepts, the definition of the information cocoon is relatively broader. It refers to the homogenization of user information caused by multiple influencing factors, which is the primary focus of this study. We aim to analyze and assess the factors that influence the information cocoon from both individual and group perspectives. By integrating these two perspectives, we seek to offer a more comprehensive understanding of the cocoon effect and its underlying causes.

\section{Assessment Indicators}
The study analyzes the effect of information cocoons from both individual and group perspectives, which are shown in Figure \ref{Fig.architecture}. %Among them, the individual perspective includes narrowing the information field of view and individual opinion polarization, while the group perspective includes network polarization and opinion polarization.

\subsection{Individual-level Indicators}
From an individual perspective, information cocoons can manifest themselves as two key aspects: the narrowing of the information scope and the polarization of opinions. 
The former refers to limited exposure to diverse sources.
%the situation where a person is exposed to information from a limited range of sources lacking diverse views. 
This can be measured through the diversity of the recommendation list, with the indicators such as the number of topic categories and information entropy. 
The latter occurs when an individual’s views or attitudes change to an extreme position, leading to rejection or hostility toward opposing viewpoints. This can be quantified by the selective clicking behavior in the recommendation list. If a user continues to select items that match their historical preferences, even when the list contains diverse information, it indicates opinion polarization, with the click repeat rate being a key indicator.

\subsubsection{Number of Topic Categories}
A fundamental measure of diversity is the number of distinct topic categories in the recommendation list. We define the average number of unique categories in users’ Top-$K$ recommendations as follows:

\begin{equation}
    \bar{\mathcal N} = \frac{1}{M} \sum_{j=1}^{M} |n_j|,
\end{equation}

\noindent where $M$ denotes the total number of users, $n_j$ is the set of unique topic categories appearing in the Top-$K$ lists of the $j$-th user. A lower $\bar{\mathcal N}$ indicates reduced topical diversity, suggesting a stronger information cocoon effect.

\subsubsection{Category Information Entropy}
Information entropy quantifies the uncertainty and diversity of topic categories within a recommendation list. Higher entropy indicates broader topical coverage, while lower entropy reflects a narrower information scope. It is formulated as:

\begin{equation}
    \bar{\mathcal H} = \frac{1}{M} \sum_{j=1}^{M} \left( -\sum_{i=1}^{n_j} p^{j}_i \log p^{j}_i \right),
\end{equation}

\noindent where $p^{j}_i$ denotes the proportion of category $i$ in the $j$-th user's Top-$K$ list. A lower $\bar{\mathcal H}$ reflects reduced topic diversity and a more pronounced information cocoon effect.

\subsubsection{Click Repeat Rate}
To assess users' tendency to select familiar content, we simulate users' click behavior and compute the proportion of clicked items whose categories overlap with users’ historical interests. A higher repeat rate shows more conservative click behavior and a stronger cocoon effect, as users repeatedly select news with similar topics rather than exploring new ones. It is defined as:

\begin{equation}
    \bar{\mathcal R} = \frac{1}{M} \sum_{j=1}^{M} \left( \frac{1}{L_j} \sum_{u=1}^{L_j} \mathbb{I}(l_u \in h_j) \right),
\end{equation}

\noindent where $M$ is the total number of users, $L_j$ is the number of clicked items by user $j$, $l_u$ is the category of the $u$-th clicked item, $h_j$ is the set of historical click categories for user $j$, and $\mathbb{I}(\cdot)$ is the indicator function. A higher $\bar{\mathcal R}$ reflects stronger topic conservatism and a deeper cocoon effect.

\subsection{Group-level Indicators}
From a group perspective, the information cocoon can be summarized in terms of group network polarization and group opinion polarization. The former refers to large-scale clustering of group members, where frequent interactions with nearby neighbors increase homogenization. This can be represented using social network characteristics, such as network density. The latter occurs when views or attitudes within the group become more extreme. It is often due to group members always reinforcing shared views through repeated like-minded interactions, while opposing perspectives are minimized or rejected. According to weak tie theory \cite{granovetter1973strength}, the weak ties between groups are crucial for external communication and exposure to new information. Strong ties within the group reinforce existing beliefs, leading to a closed system. Thus we quantify the difference between internal (within-community) and external (between-community) connections to measure the persuasive influence of in-group versus out-group opinions, reflecting the community's openness to diverse views.

Before defining specific indicators, we model the user-news interaction network as a bipartite graph $\mathcal{G} = (\mathcal{U} \cup \mathcal{N}, \mathcal{E})$. $\mathcal{U} = \{u_1, u_2, \dots, u_{n_{\text{user}}}\}$ represents user nodes and $\mathcal{N} = \{n_1, n_2, \dots, n_{n_{\text{news}}}\}$ represents news nodes. $\mathcal{E} \subseteq \mathcal{U} \times \mathcal{N}$ denotes the set of observed user-news interactions.

\subsubsection{Network Density}
Network density primarily refers to the scale and cohesiveness of homogeneous network groups, reflecting the aggregation and consistency within the group. We perform community detection over $\mathcal{G}$ and compute the average internal density across all communities.

\begin{equation}
    \bar{\mathcal D} = \frac{1}{C} \sum_{c=1}^{C} \frac{|\mathcal{E}^{c}_{\text{in}}|}{|\mathcal{U}^{c}| \cdot |\mathcal{N}^{c}|}.
\end{equation}

\noindent For each community $c \in \{1, 2, \dots, C\}$, $\mathcal{U}^{c} \subseteq \mathcal{U}$ and $\mathcal{N}^{c} \subseteq \mathcal{N}$ denote the users and news nodes in $c$. $\mathcal{E}^{c}_{\text{in}} \subseteq \mathcal{U}^{c} \times \mathcal{N}^{c}$ denotes the internal edges of $c$. A higher $\bar{\mathcal D}$ indicates stronger intra-community interaction density, reflecting tighter clustering and a more pronounced information cocoon effect.

\subsubsection{Community Openness}
The internal and external edges of a community directly reflect its communication status. We measure the weighted difference between internal and external edges to capture changes in community openness.

\begin{equation}
    \bar{\mathcal O} = \frac{1}{C} \sum_{c=1}^{C} \frac{|\mathcal{E}^{c}_{\text{out}}| - |\mathcal{E}^{c}_{\text{in}}|}{|\mathcal{E}^{c}_{\text{out}}| + |\mathcal{E}^{c}_{\text{in}}|}.
\end{equation}

\noindent For each community $c \in \{1,2, \dots, C\}$, $\mathcal{E}^{c}_{\text{out}}$ is the set of external edges. A higher $\bar{\mathcal O}$ indicates greater inter-community exposure, while a lower or negative value suggests increased insularity and a more severe information cocoon effect.

\begin{table*}[t]
\centering
    \caption{Statistics of the datasets.}
    \label{tab:datasets}
    \begin{tabular}{c | c  c  c  c c c c}
    \toprule
Dataset & News & Users & Category & Subcategory & Impression &Behavior &language\\
\cmidrule{1-8}
Adressa& 923 & 15514 & 19 & 103 & 46,542 & 2,717,915&Norwegian\\
MIND & 161,013 & 1,000,000 & 20 & 257 & 15,777,377 & 24,155,470&English\\
\bottomrule
\end{tabular}
\end{table*}

\section{Evaluation}
\subsection{Experimental Settings}
 
\subsubsection{Datasets}
We used two widely adopted news datasets, MIND \cite{wu-etal-2020-mind} and Adressa \cite{gulla2017adressa}, as the basis of our experiments to ensure the credibility and representativeness of the results, as shown in Table \ref{tab:datasets}. The first is the Adressa\footnote{https://reclab.idi.ntnu.no/dataset/} \cite{gulla2017adressa} dataset, jointly released by Norwegian news publishers such as Adresseavisen and the Norwegian University of Science and Technology, collected from the website www.adresseavisen.no. The second is a large-scale, real-world dataset—the Microsoft News Dataset (MIND\footnote{https://msnews.github.io/}) \cite{wu-etal-2020-mind}, which is constructed from user click logs on Microsoft News and serves as a benchmark for news recommendation research.
% The detailed information is shown in Appendix.

\subsubsection{News Recommenders}
We use several typical news recommendation models to examine the performance of information cocoons in our experiments, including NRMS \cite{wu-2019-NRMS}, NAML \cite{Wu2019NAML}, LSTUR \cite{an-etal-2019-LSTUR}, DKN \cite{Wang-2018-DKN}, TANR \cite{wu-etal-2019-TANR}, NPA \cite{Wu-2019-NPA}, and Hi-Fi Ark \cite{Liu-2019-HF}. More details are displayed in Appendix \ref{News Recommenders}.

\subsubsection{User-Item Network}

Unlike social media, news services typically lack explicit social connections such as follower relationships. To examine group-level information cocoons, we construct a user-item bipartite network based on click behaviors \cite{li2023breaking}, where nodes denote users and news, and edges denote interactions. This graph captures users' preferences and their structural distribution. To better understand the user-news and user-user relationships, we apply community detection to this graph. Each community corresponds to a cluster of users with similar interests and the news they consume. Fewer and larger communities indicate higher homogeneity, while sparser inter-community links suggest stronger polarization. These features serve as indicators of collective information cocooning. We adopt the Louvain algorithm \cite{Blondel2008FastUO} for community detection, which optimizes modularity by assigning nodes to communities with the greatest local modularity gain. This method effectively identifies groups with dense internal and sparse external links, allowing us to analyze the relationship between information cocooning and community structure.

\subsubsection{Hyper-parameter Settings}
We implement the experiments based on PyTorch and use the Adam optimizer. The learning rate is set to $10^{-4}$, and the dropout rate is set to 0.2. For each user, we sample up to 50 news articles from their click history. The maximum number of words for the title and abstract is set to 20 and 50, respectively.

\subsection{Overall Results Performance}
Based on the experimental settings, we conducted information cocoon assessment and analysis across \textbf{multiple rounds} of recommendations using different news recommendation models, as shown in Table \ref{tab:results}. More detailed results and specific analysis are available in the Appendix \ref{Detailed Experimental Results}.

\begin{table*}
\centering
\caption{Results of information cocoon metrics in multi-round recommendations.}
    \label{tab:results}
\begin{tabular}{ c |c c c c c ||c| c c c c c }
\toprule
\textbf{MIND} & N@20 & H@20 & R & D & O &\textbf{Adressa} &N@20 & H@20 & R & D & O \\
\cmidrule{1-12}
NAML  & \textbf{4.5611} & \textbf{1.2511} & 0.9822 & 0.0011 & \textbf{0.1502} &NAML & \textbf{4.0267} & \textbf{0.9780} & \underline{0.9704} & \underline{0.0023} & \textbf{0.2253}\\
TANR  & 4.6748 & \underline{1.6419} & 0.9818 & 0.0012 & 0.2267 &TANR & 5.0238 & 1.9009 & 0.9690 & 0.0014 & 0.6467 \\
NRMS  & 5.0536 & 1.6676 & \underline{0.9845} & \textbf{0.0015} & 0.3428 &NRMS & 4.9461 & 1.8959 & 0.9691 & 0.0020 & 0.6139 \\
Hi-Fi Ark  & \underline{4.6357} & 1.6944 & \textbf{0.9859} & 0.0013 & 0.5863 &Hi-Fi Ark & 5.0470 & 1.9464 & 0.9488 & 0.0013 & 0.6612 \\
NPA  & 5.1058 & 1.6891 & 0.9831 & \underline{0.0014} & \underline{0.1821} &NPA & 5.0398 & 1.8941 & \textbf{0.9789} & 0.0016 & 0.6417\\
LSTUR  & 6.0568 & 2.7855 & 0.9600 & 0.0013 & 0.6896 &LSTUR & \underline{4.2220} & \underline{1.2548} & 0.9699 & \textbf{0.0023} & \underline{0.5679} \\
DKN  & 7.3984 & 2.8459 & 0.9332 & 0.0005 & 0.9293 &DKN & 5.1290 & 1.9751 & 0.9388 & 0.0015 & 0.8202 \\
\bottomrule
\end{tabular}
\end{table*}

% Figure 3
\begin{figure}[t]
\centering
    \subfigure[\textbf{Category-Before}]{
\includegraphics[width=0.47\linewidth]{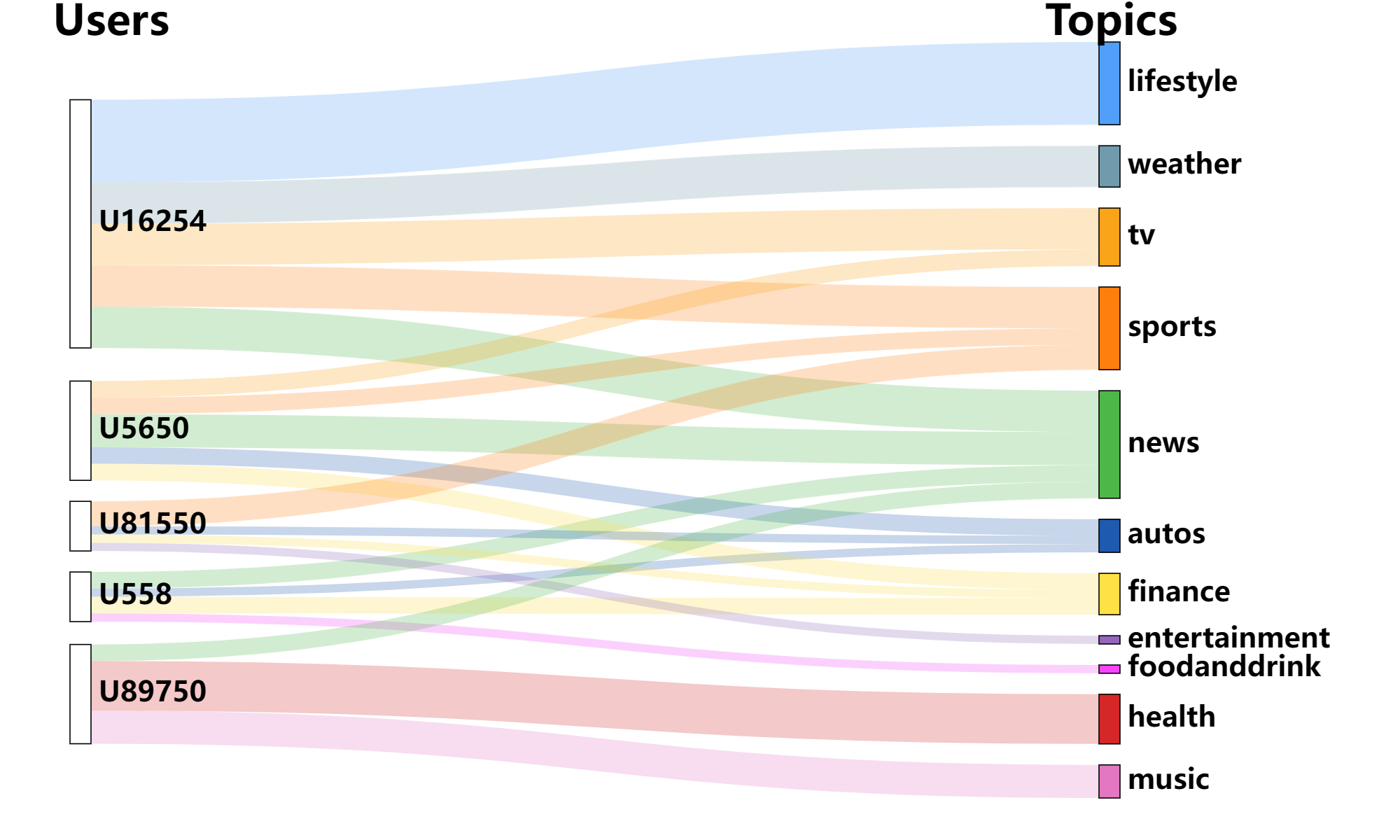}
}
    \subfigure[\textbf{Category-After}]{
\includegraphics[width=0.47\linewidth]{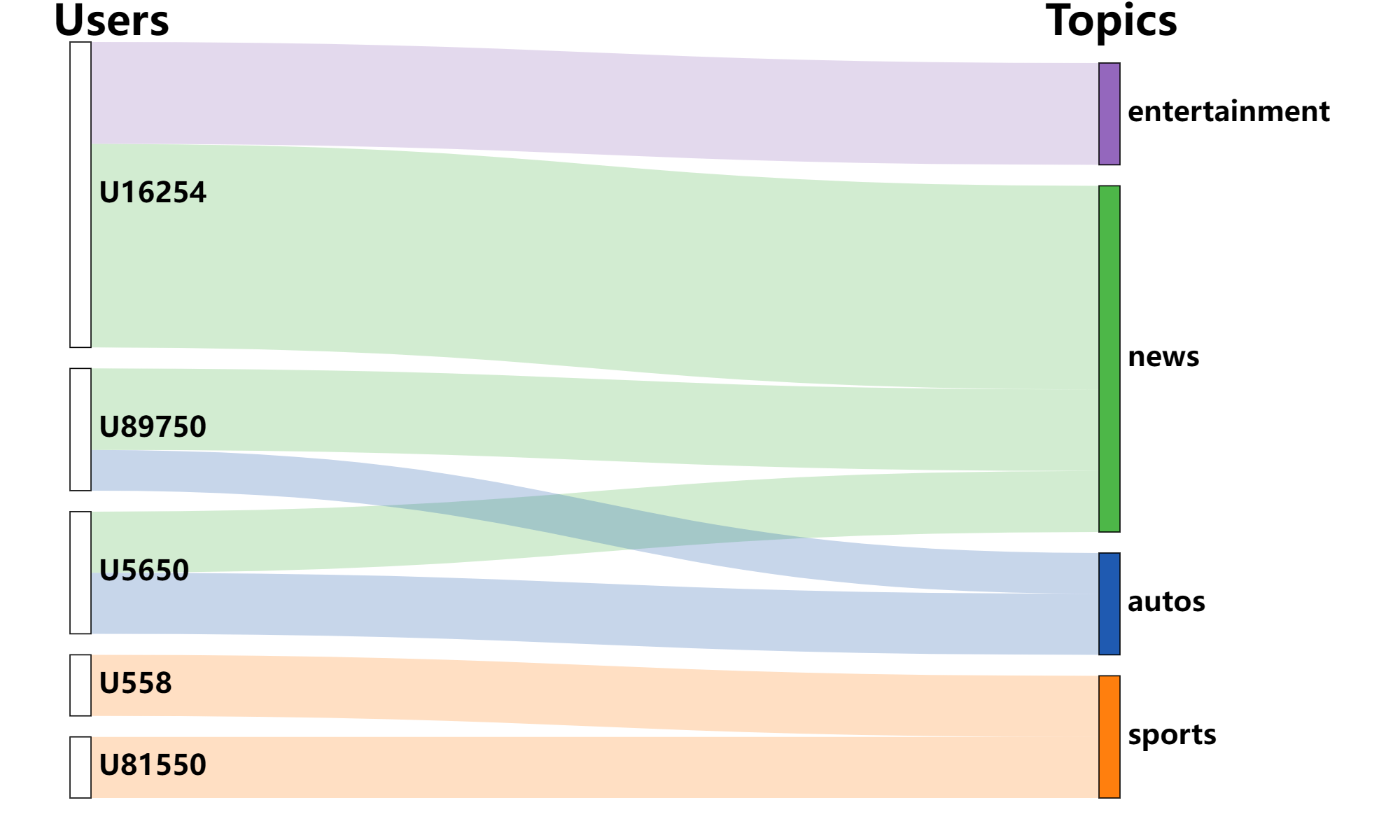}
}
    \subfigure[\textbf{Subcategory-Before}]{
\includegraphics[width=0.47\linewidth]{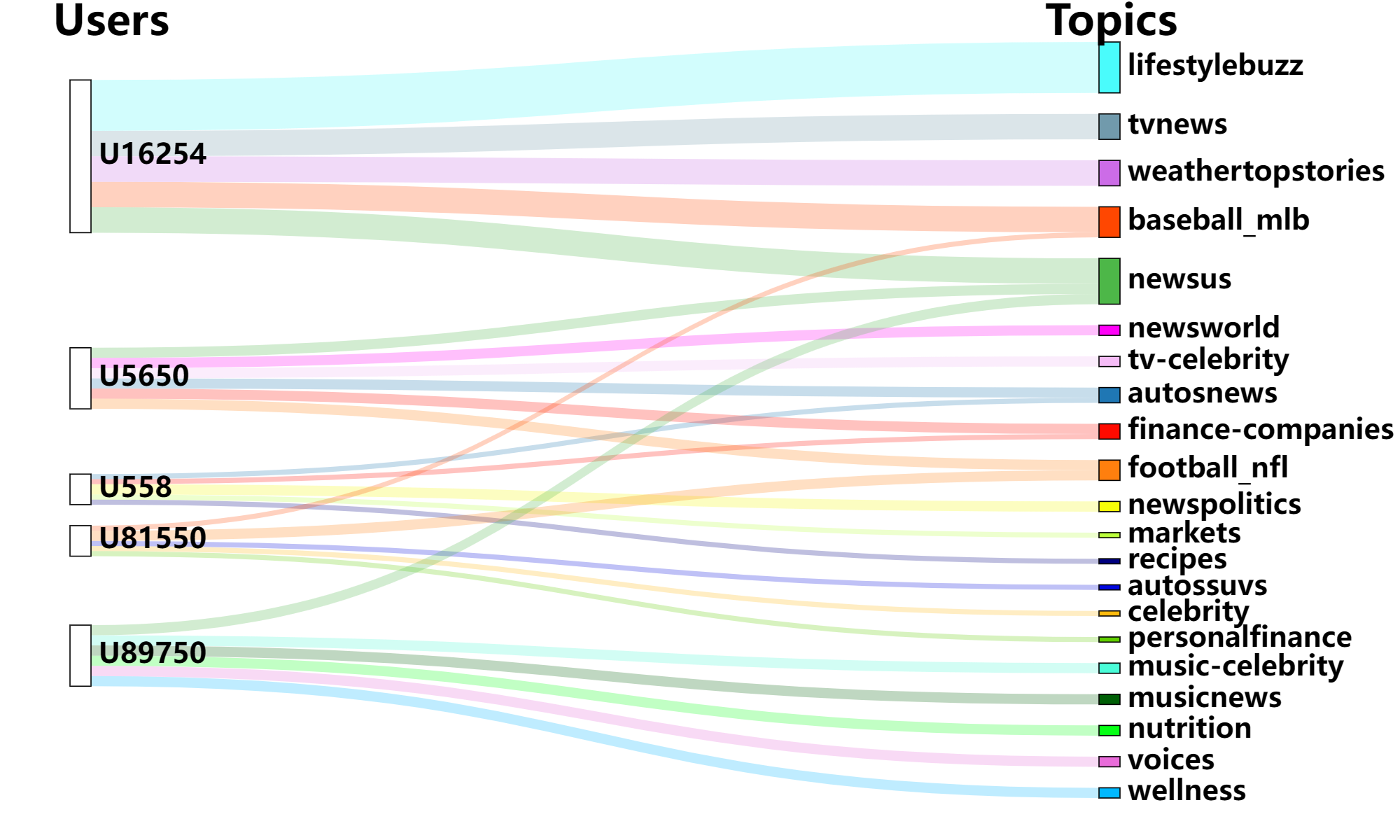}
}
    \subfigure[\textbf{Subcategory-After}]{
\includegraphics[width=0.47\linewidth]{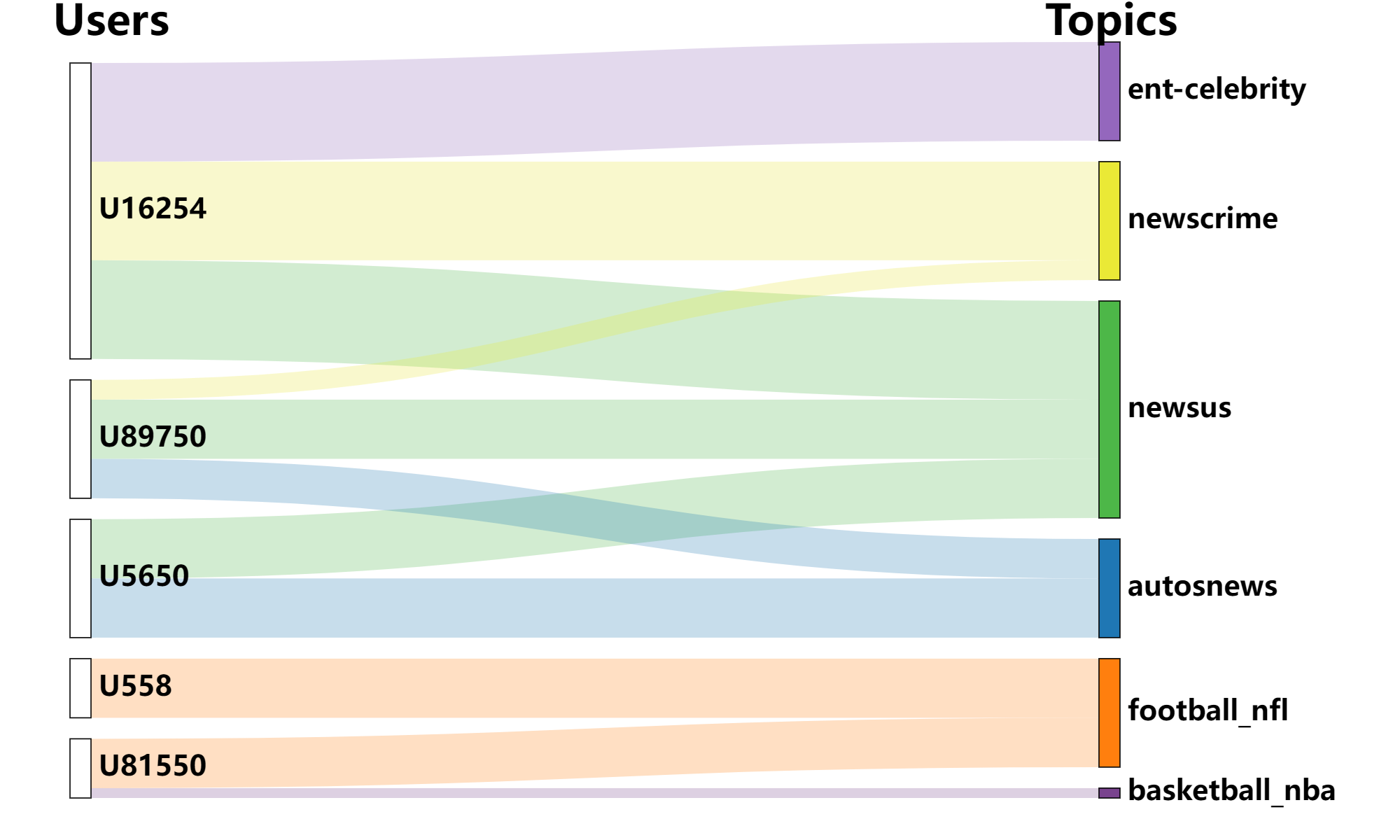}
}
\caption{The Sankey diagram compares the same five users' original history with the Top-6 list topic categories after multiple recommendation rounds. The results of the \textit{NRMS} model on the MIND dataset are used here as an example.}
\label{Fig.Homogenization Results}
\end{figure}

% Figure 4
\begin{figure}[ht]
\centering
\subfigure[\textbf{Before Recommendation}]{
\includegraphics[width=0.47\linewidth]{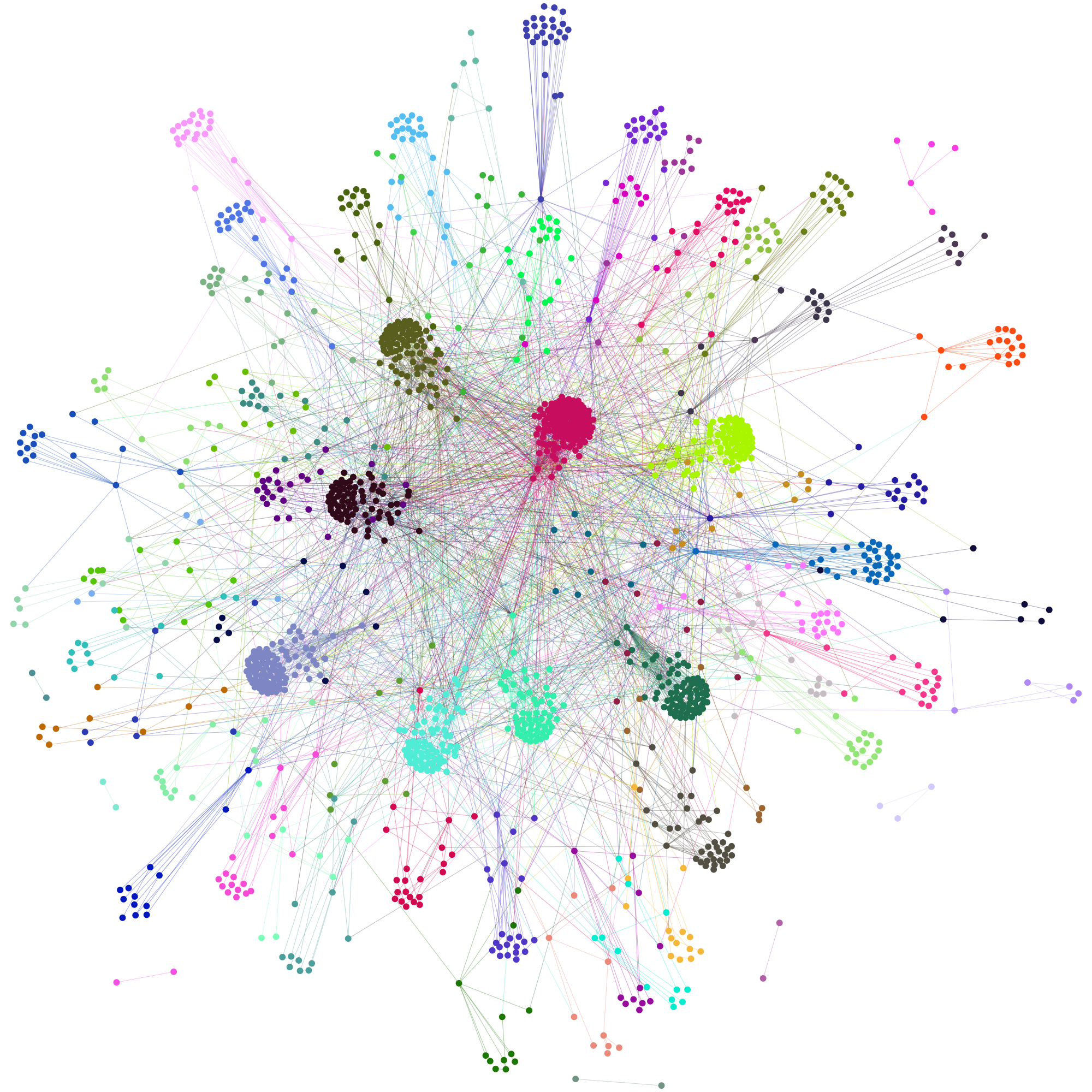}
}
\subfigure[\textbf{After Recommendation}]{
\includegraphics[width=0.47\linewidth]{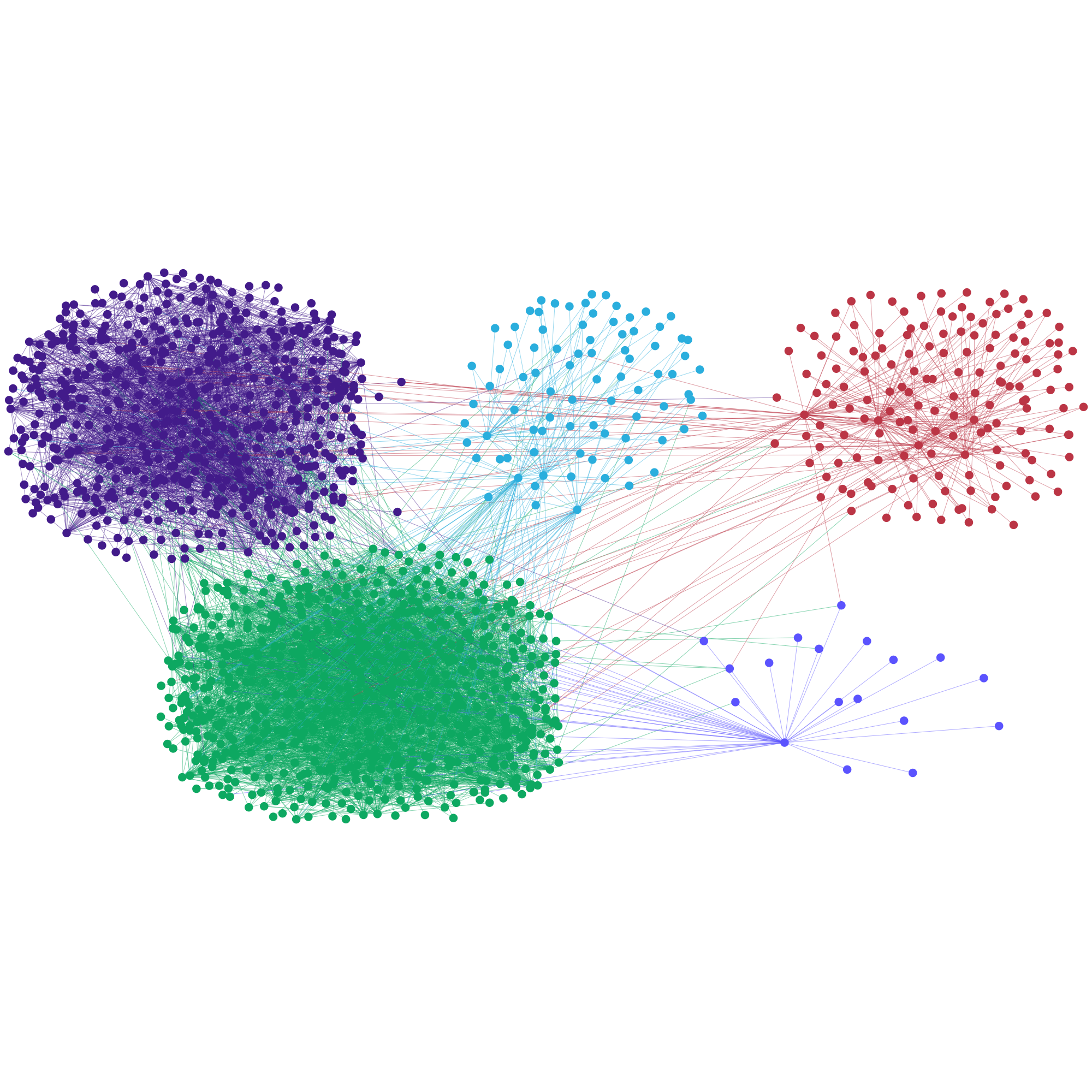}
}
\caption{The User-Item networks are based on click history and its evolution across multiple recommendation rounds. Using the Louvain algorithm, nodes are grouped into communities for different colors. It uses the results of the \textit{NRMS} model and includes 6000 randomly selected users.}
\label{Fig.Polarization Results}
\end{figure}

From both individual and group perspectives, most models reflect the gradual deepening of the cocoon effect during recommendations, including a reduction in the topic diversity of recommendation lists and the densification and closure of the user-item network.
Using the \textit{NRMS} model results shown in Figure \ref{Fig.Homogenization Results} as an example, we can observe a significant decrease in the diversity of news topics in users’ click lists after multiple rounds of recommendation, compared to their historical click lists, which is evident for both category and subcategory. Additionally, the user-item network shown in Figure \ref{Fig.Polarization Results} undergoes significant changes, with the community distribution shifting from diverse and open to more aggregated and closed. These findings highlight the deepening cocoon effect in recommendations, where users are increasingly confined to homogeneous information.

%Figure 5
\begin{figure*}[t]
\centering
\subfigure[\textbf{MIND-Category@20}]{
\includegraphics[width=5.5cm,height=4.5cm]{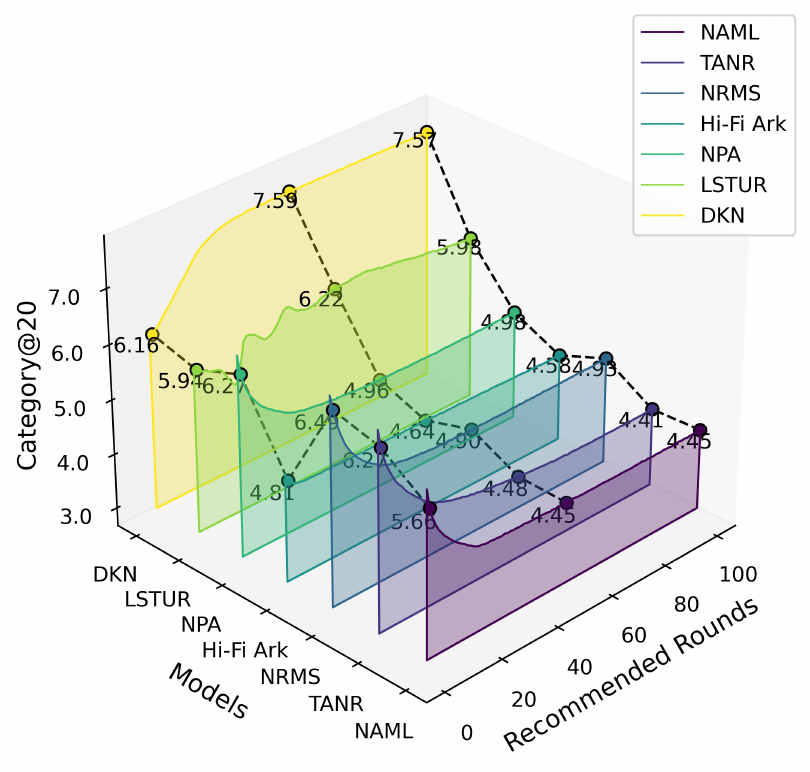}
\label{category20_category}
}
\subfigure[\textbf{MIND-Entropy@20}]{
\includegraphics[width=5.5cm,height=4.5cm]{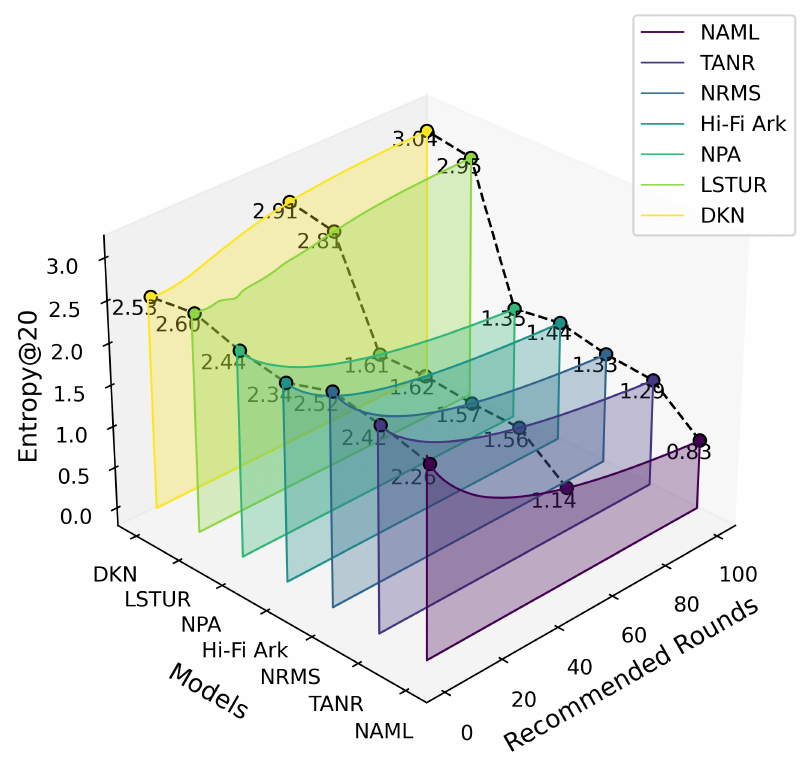}
\label{entropy20_category}
}
\subfigure[\textbf{MIND-Click Repeat Rate}]{
\includegraphics[width=5.5cm,height=4.5cm]{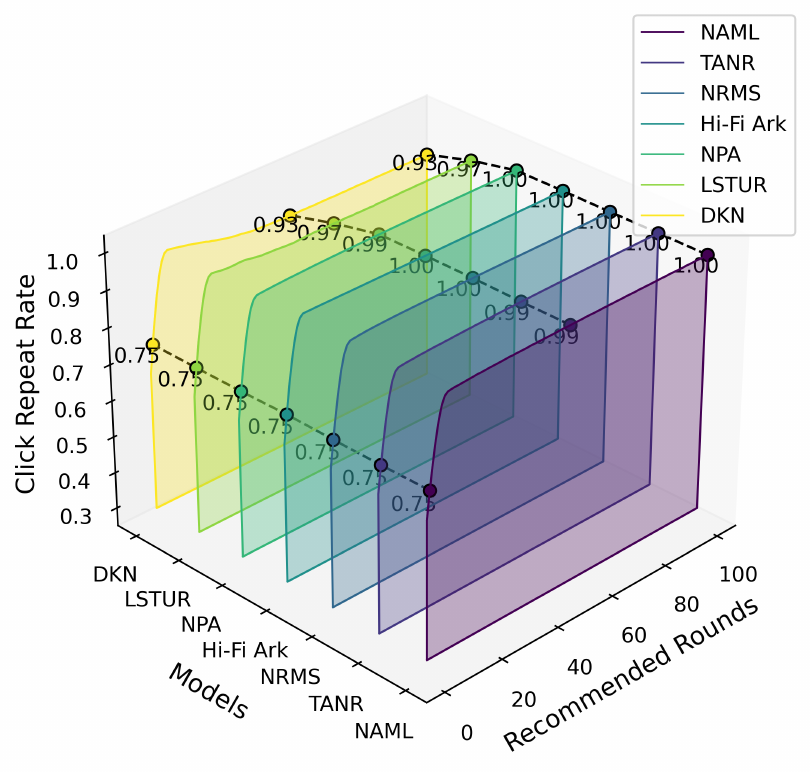}
\label{click_category}
}
\subfigure[\textbf{Adressa-Category@20}]{
\includegraphics[width=5.5cm,height=4.5cm]{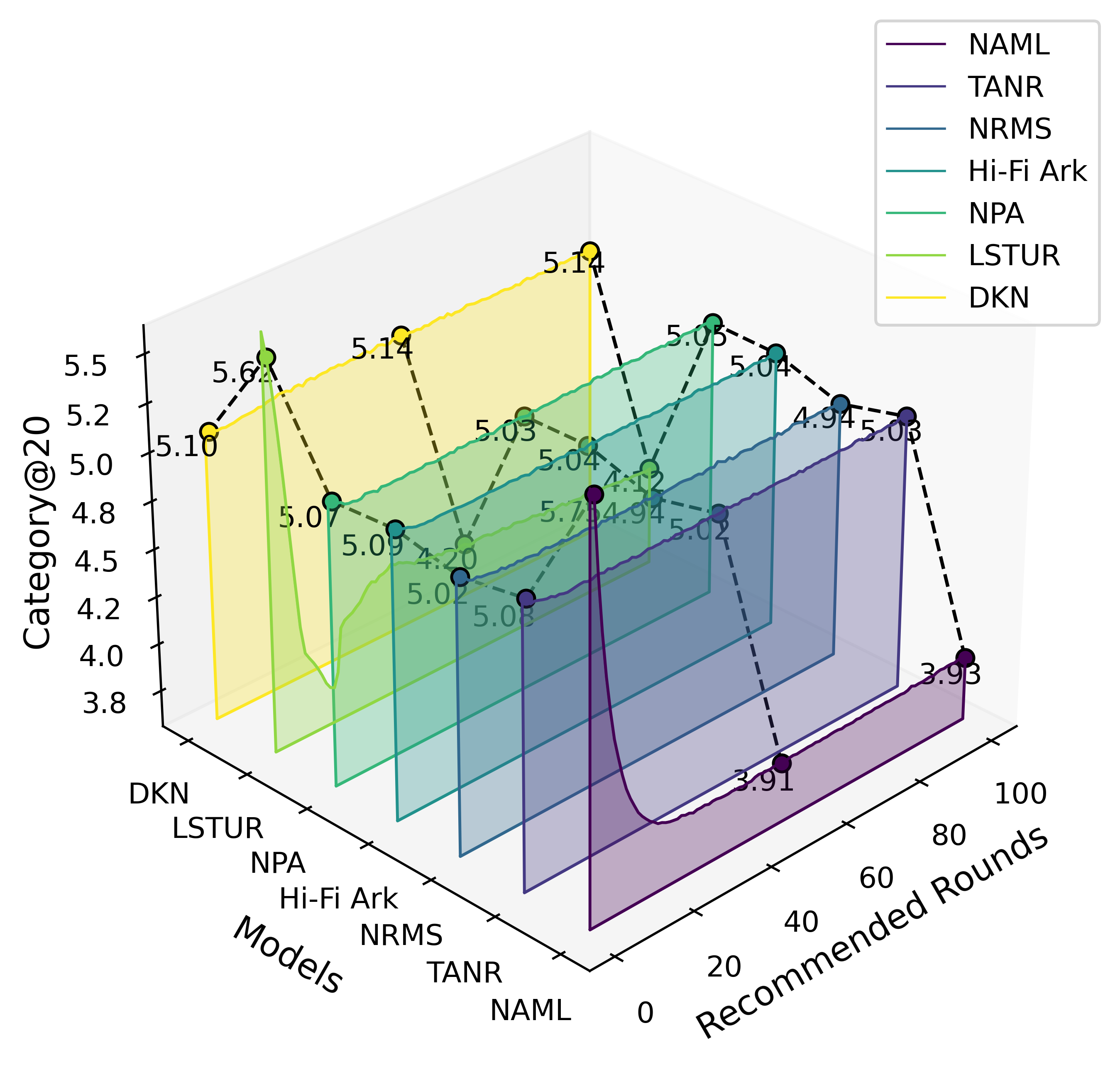}
\label{category20_subcategory}
}
\subfigure[\textbf{Adressa-Entropy@20}]{
\includegraphics[width=5.5cm,height=4.5cm]{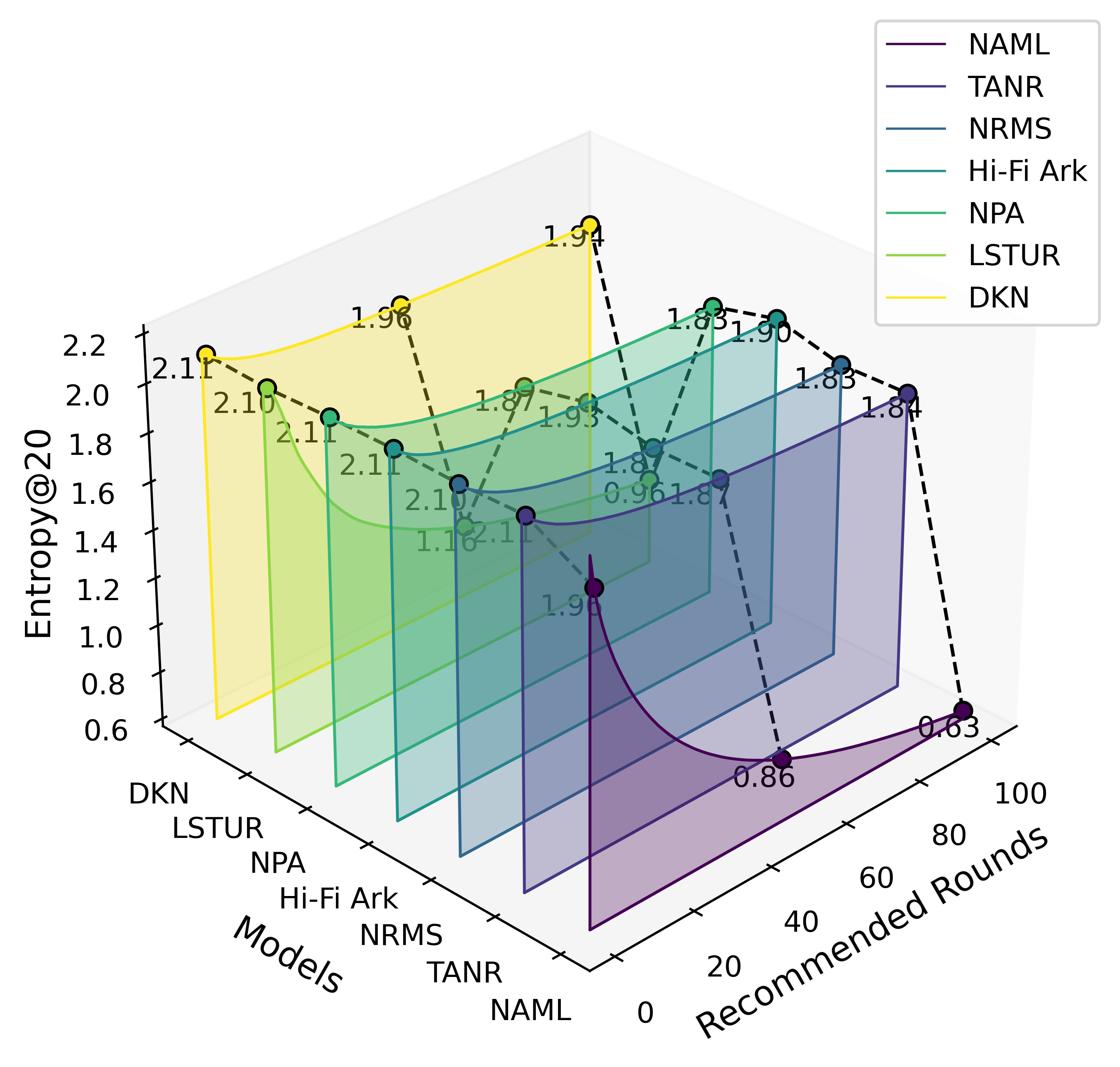}
\label{entropy20_subcategory}
}
\subfigure[\textbf{Adressa-Click Repeat Rate}]{
\includegraphics[width=5.5cm,height=4.5cm]{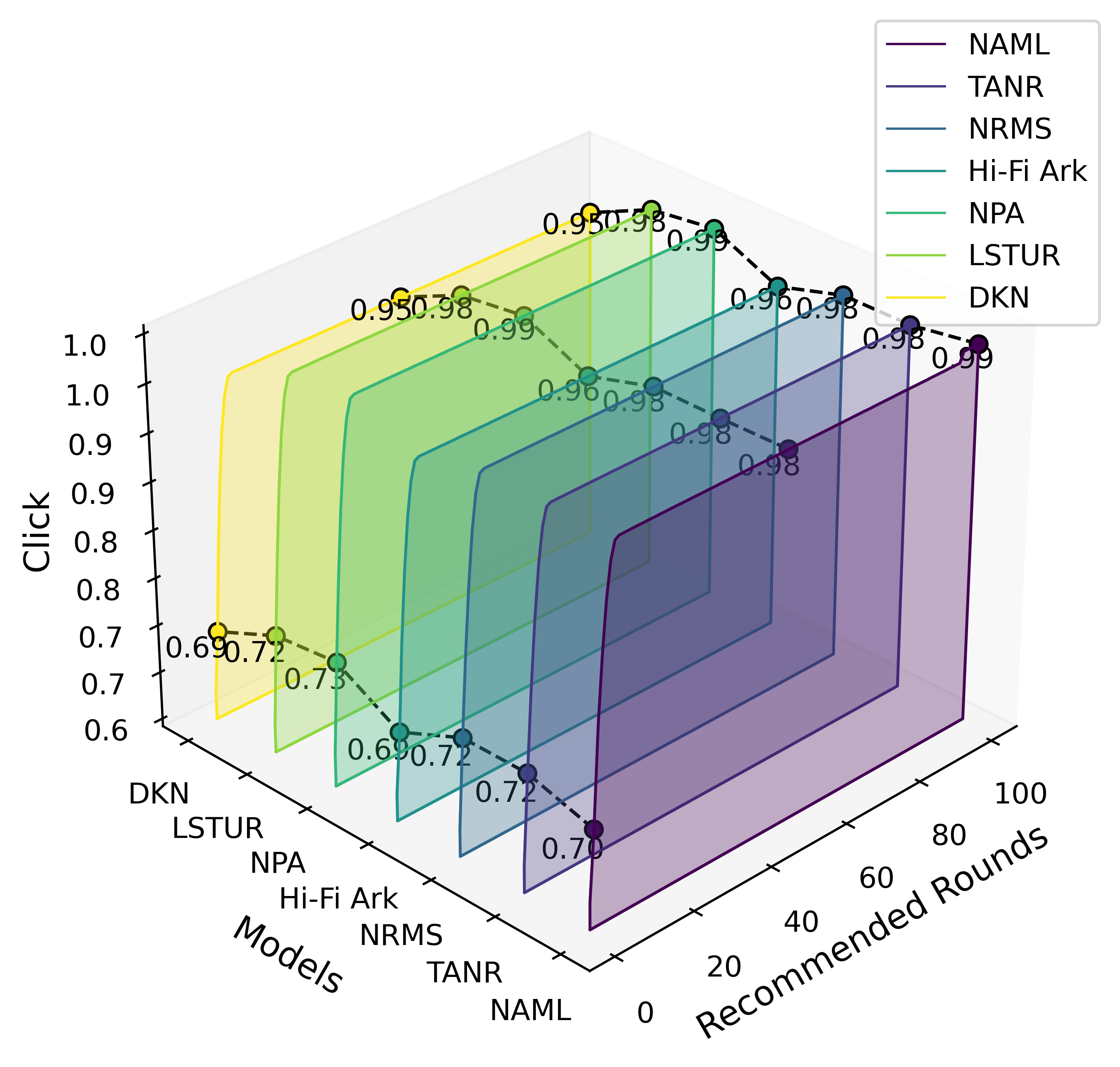}
\label{click_subcategory}
}
\caption{Number of topic categories and category information entropy of the Top-20 list, and click repeat rate after multiple recommendation rounds for each model under the individual perspective.}
\label{Fig.individual}
\end{figure*}

\begin{table}
\centering
\caption{Comparison results of categories and subcategories}
  \label{tab:categories and subcategories}
\begin{tabular}{ c  |c c c c c }
\toprule
\textbf{MIND} & N@20& H@20& R& D& O\\
\cmidrule{1-6}
NAML & \textbf{4.5611} & \textbf{1.2511} & 0.9822 & 0.0011 & \textbf{0.1502} \\
 & \textbf{9.1109}& \textbf{2.5144}& 0.8897& 0.0011& \textbf{0.2359}\\
 \cmidrule{1-6}
TANR  & 4.6748 & \underline{1.6419} & 0.9818 & 0.0012 & 0.2267 \\
 & 9.9821& 3.3832& 0.8642& 0.0011& 0.3153\\
 \cmidrule{1-6}
NRMS  & 5.0536 & 1.6676 & \underline{0.9845} & \textbf{0.0015} & 0.3428 \\
 & \underline{9.8232}& \underline{3.2295} & \textbf{0.9103}& \textbf{0.0014}& 0.4096\\
 \cmidrule{1-6}
Hi-Fi Ark  & \underline{4.6357} & 1.6944 & \textbf{0.9859} & 0.0013 & 0.5863 \\
 & 10.4033& 3.4541& \underline{0.9021} & 0.0014& 0.6171\\
 \cmidrule{1-6}
NPA  & 5.1058 & 1.6891 & 0.9831 & \underline{0.0014} & \underline{0.1821} \\
 & 10.0241& 3.3190& 0.8876& \underline{0.0014} & \underline{0.2633}\\
 \cmidrule{1-6}
LSTUR  & 6.0568 & 2.7855 & 0.9600& 0.0013 & 0.6896 \\
 & 10.2787& 4.4151& 0.6382& 0.0013& 0.7413\\
 \cmidrule{1-6}
DKN  & 7.3984 & 2.8459 & 0.9332 & 0.0005 & 0.9293 \\
 & 13.5814& 4.6780& 0.6738& 0.0055& 0.9478\\
 \bottomrule
\end{tabular}
\end{table}

\subsection{Individual Homogenization} 
\label{Homogenization of Recommendations}
Figure \ref{Fig.individual} presents the results of the recommendation algorithms on three individual-level assessment metrics. As can be seen from the figure, for the majority of models, the number of topic categories and category information entropy decrease, while the click repeat rate increases. This indicates a growing similarity in recommended content over time, reflecting a deepening information cocoon effect. The cocoon tends to intensify more significantly on the larger dataset MIND over multiple rounds of recommendation.

The results on categories and subcategories shown in Table \ref{tab:categories and subcategories} indicate that most recommendation algorithms exhibit consistent trends across both levels of granularity. Models with a stronger cocoon effect at the category level tend to show similarly pronounced effect at the subcategory level. However, since subcategories are more numerous and finer-grained, the resulting metrics typically reflect a milder degree of cocooning. It suggests that although the general trend holds, the categories can more clearly reflect the degree of cocoon. More details will be shown in the Appendix \ref{The comparison of categories and subcategories}.

\subsection{Group Polarization} \label{Social Polarization}

\begin{figure}[t]
\centering
\subfigure[\textbf{MIND}]{
\includegraphics[width=0.47\linewidth]{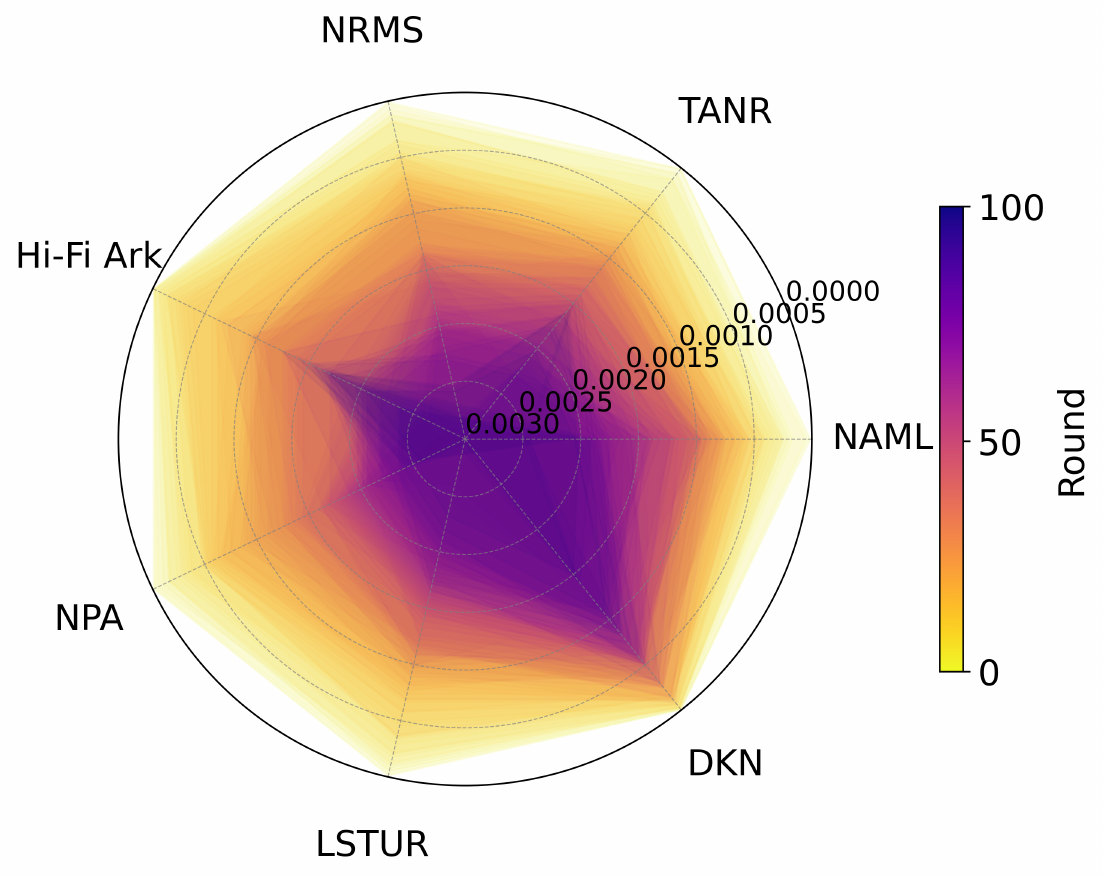}
}
\subfigure[\textbf{Adressa}]{
\includegraphics[width=0.47\linewidth]{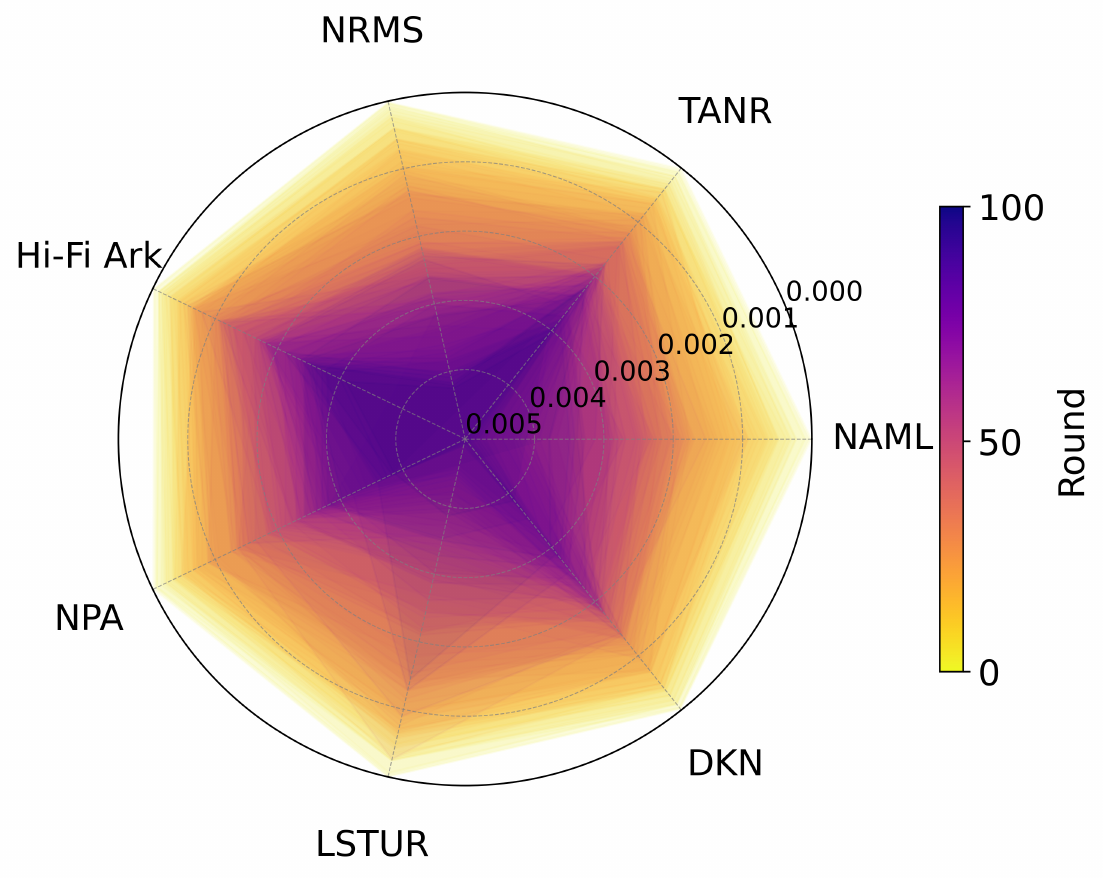}
}
\caption{Network density after multiple recommendation rounds for each model under the group perspective.}
\label{Fig.density}
\end{figure}

\begin{figure}[t]
\centering
\subfigure[\textbf{MIND}]{
\includegraphics[width=0.47\linewidth]{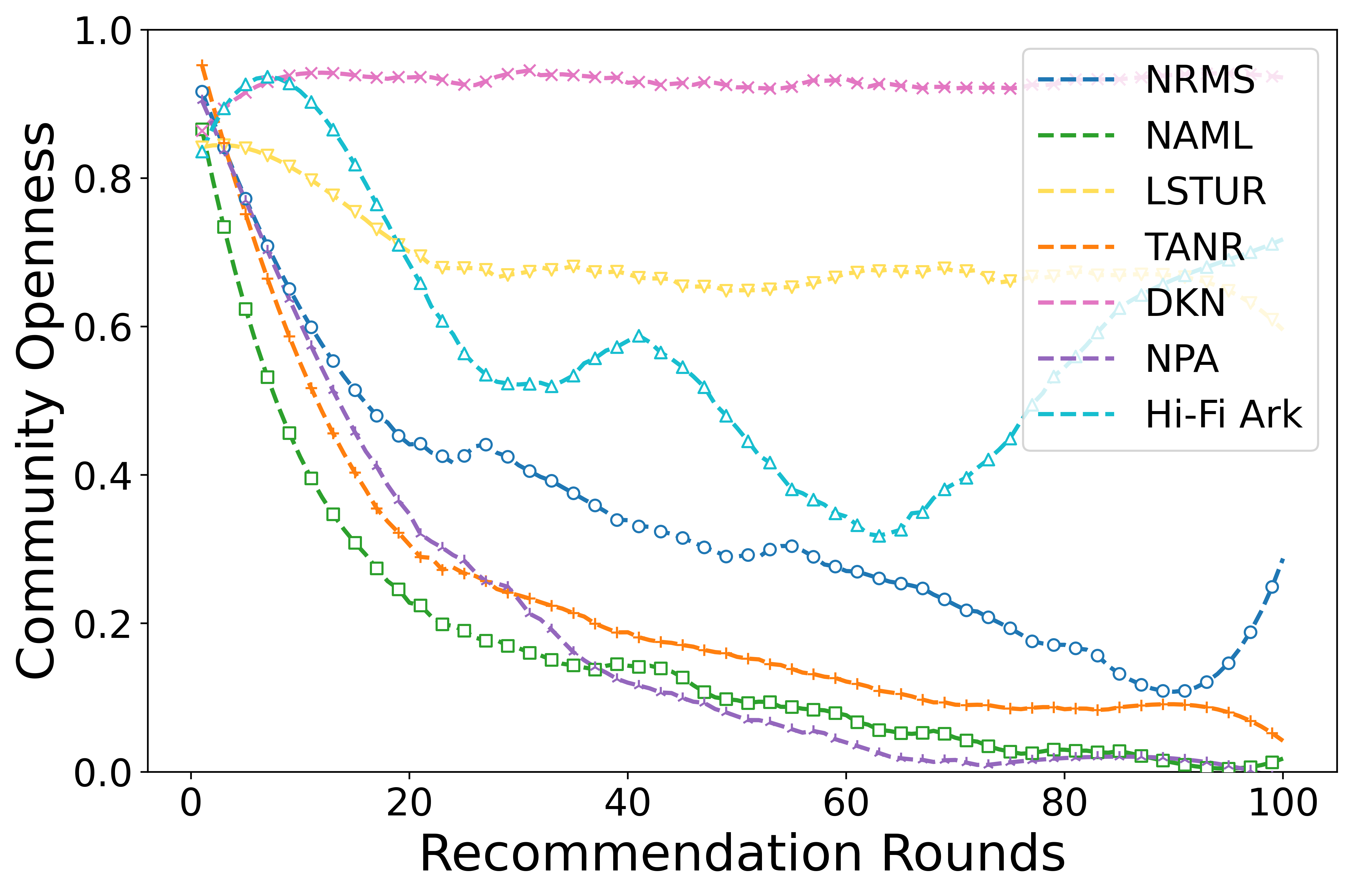}
}
\subfigure[\textbf{Adressa}]{
\includegraphics[width=0.47\linewidth]{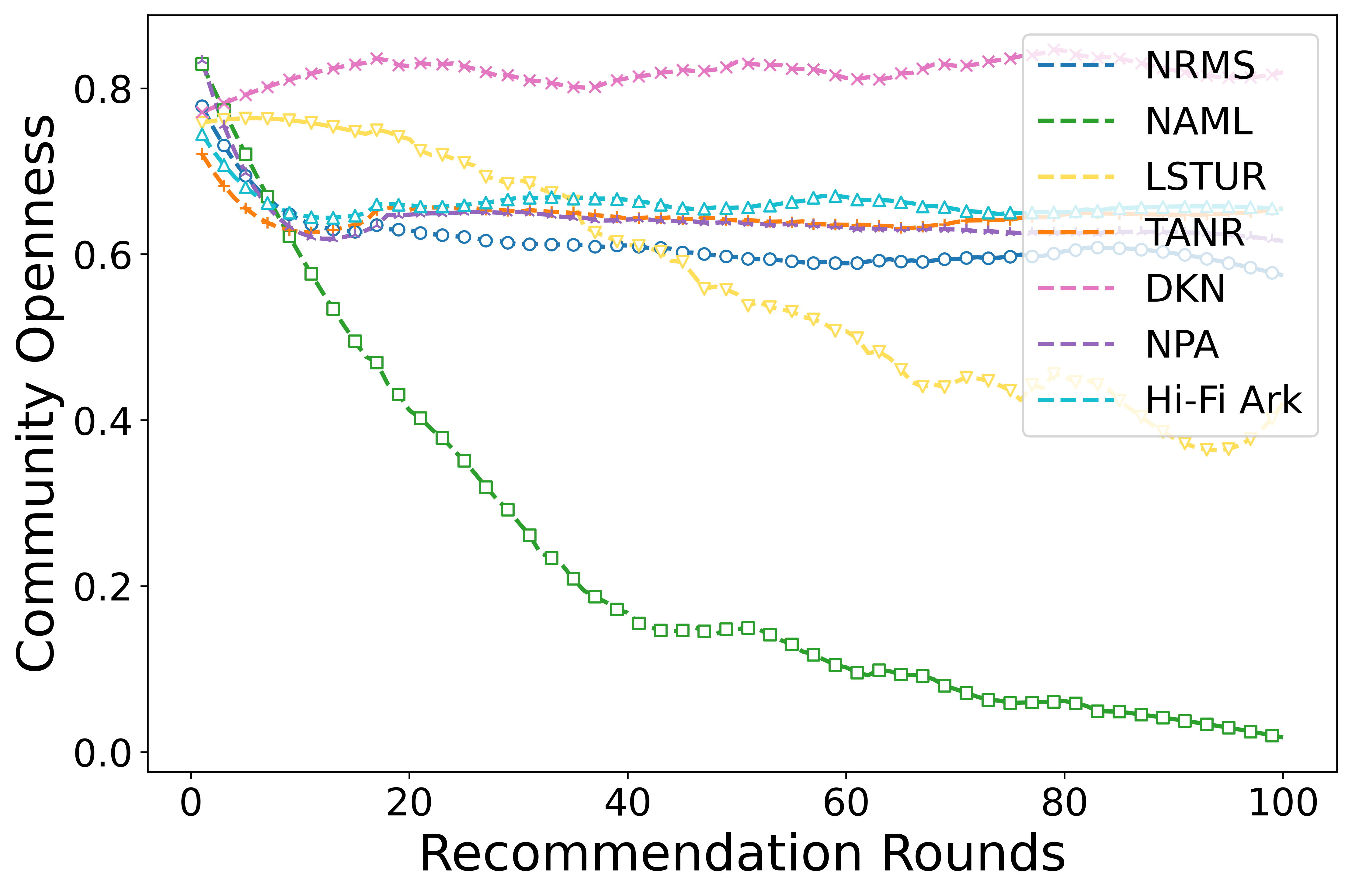}
}
\caption{Community openness after multiple recommendation rounds for each model under the group perspective.}
\label{Fig.openness}
\end{figure}

The results of the group polarization metrics are shown in Figure \ref{Fig.density} and Figure \ref{Fig.openness}. For most models, network density increases and community openness decreases, indicating that the recommended news becomes more concentrated within communities, reducing the exposure to outside content. The user-item network gradually forms a more closed and clustered community, suggesting growing group polarization and a deepening information cocoon effect. It also intensifies more significantly on the larger dataset MIND over multiple rounds of recommendation.
%that the degree of group polarization strengthens, which also signifies an intensification of the information cocoon.

Similar to the individual level, group-level metrics across different news recommendation models exhibit consistent patterns when evaluated using both categories and subcategories. Although category-based metrics still tend to indicate slightly stronger cocoon effects than those based on subcategories, the differences are marginal. This stability may arise because group-level metrics are derived from the global structural properties of the user-item network and are less directly influenced by the number of topics.

\subsection{Recommendation Algorithms Comparison} \label{Influence of Recommendation Algorithms}
Based on the above results, we analyze the impact of different recommendation algorithms on the performance of the information cocoon metrics. It is evident that most news recommendation models exacerbate the information cocoon effect across the indicators. However, different models exhibit different degrees of information cocoon deepening. There are also some models that maintain a certain degree of diversity. 
%For example, the diversity of recommendation topics of LSTUR and DKN does not decrease significantly. But we find that this may mostly stem from the fact that their topic categories are already more homogeneous at the time of initial recommendation.

At the individual level, the information cocoon effect is most pronounced in the \textit{NAML} model. In contrast, models such as \textit{DKN} exhibit less significant changes in the number of topic categories and category information entropy. \textit{LSTUR} shows differences across two datasets of different scales. The detailed analyses are as follows: (1) Both \textit{NRMS} and \textit{NAML} utilize attention mechanisms, with \textit{NAML} additionally treating the news title, content, and topic categories as distinct views. This precise modeling approach makes the recommender highly focused on the user's known interests, leading to recommendations that increasingly center on specific categories. (2) \textit{DKN} integrates external knowledge graphs, incorporating richer semantic information into the recommendation process. The knowledge-driven mechanism allows the system to consider not only users' behaviors but also the semantic relationships between content items, which increases content diversity. (3) \textit{LSTUR} models both long-term and short-term user interests in a hierarchical manner, which helps avoid the over-concentration on any particular types of content and maintain a certain level of diversity in recommendations. But when applied to the small dataset Adressa which only includes short-term data within a week, its emphasis on long-term preferences is no longer effective; instead, the echo chamber effect is significant.

% \item \textit{TANR}, which especially emphasizes temporal information, can capture the evolution of user interests over time. As time progresses, the system becomes more precise in recommending content aligned with the user's historical preferences, further reinforcing long-term interests and thereby intensifying the information cocoon. 
% \item \textit{Hi-Fi Ark} also incorporates external knowledge sources into the recommendation system, considering the semantic relationships and contextual information. But it does not show the slight increase observed in \textit{DKN}, as it prioritizes high-fidelity recommendations that closely match the user’s true needs and interests, which somewhat limits the recommendation diversity. 
% \item We also conduct a statistical analysis of the initial metric values for the \textit{LSTUR}, \textit{DKN}, and \textit{Hi-Fi Ark} models, and find that most initial results (from the first recommendation round) already exhibit a pronounced information cocoon effect.

At the group level, similar to the individual-level results, \textit{NRMS}, \textit{NAML} and \textit{NPA} exhibit the most pronounced information cocoon effect. In contrast, \textit{DKN}, which introduces external knowledge, does not show significant changes in community openness. Detailed analyses are as follows: (1) \textit{NPA} also employs a personalized attention mechanism, dynamically assigning attention based on the relevance between news items and user preferences, leading to more customized recommendations. (2) The results of \textit{DKN} suggest that the incorporation of external knowledge helps to break down barriers between communities, alleviating group polarization. (3) Similar to the individual level, the community openness of \textit{LSTUR} also declines at a slower rate in MIND, possibly due to the stability of long-term interests. However, in the short-cycle dataset Adressa, it also exhibits the deep information cocoon effect.
%As \textit{LSTUR} continues to recommend content that is more relevant to the user’s long-term interests, the system is able to maintain a degree of exploratory recommendations even in later rounds, resulting in a slower decline in community openness.

% Figure 6

% Figure 7

\subsection{Mitigation Strategies}
Based on the above evaluation and analysis results, we designed several strategies to mitigate the information cocoon effect across multiple models from the perspectives of processing and post-processing.

\textbf{Epsilon-Greedy Strategy (EGS)} introduces a controlled degree of random exploration to break feedback loops and promote exposure diversity. With a small probability, items are randomly selected from the candidate pool rather than relying solely on top-ranked predictions. It is formulated as:
%, which helps mitigate narrow personalization and supports the discovery of diverse information

\begin{equation}
P(i \mid u) = 
\begin{cases}
\frac{1}{|\mathcal{C}_u|}, & \varepsilon \\
\text{softmax}(s_{ui}), & 1 - \varepsilon
\end{cases}
\end{equation}

\noindent where $\mathcal{C}_u$ denotes the candidate item set for user $u$, $s_{ui}$ is the predicted relevance score of item $i$ for user $u$, $\varepsilon \in (0, 1)$ is the exploration probability, $P(i \mid u)$ is the final recommendation probability for item $i$.

\textbf{Content Diversity Regularization (CDR)} is a loss-level enhancement technique that penalizes semantic redundancy among recommended items. By introducing a regularization term that discourages high content similarity within recommendation lists, the model is guided to produce outputs that span a broader range of topics or styles. It is formulated as:

\begin{equation}
\mathcal{L}_{\text{CDR}} = \lambda \cdot \sum_{i, j \in \mathcal{R}_u, i \neq j} \text{sim}(\mathbf{e}_i, \mathbf{e}_j)
\end{equation}

\noindent where $\lambda$ is a hyperparameter controlling the strength of regularization, $\mathcal{R}_u$ denotes the top-$K$ recommendation list for user $u$, $\mathbf{e}_i, \mathbf{e}_j$ are the feature embeddings of items $i$ and $j$, $\mathrm{sim}(\cdot, \cdot)$ denotes the similarity function.

\textbf{Long-Term Attention Optimization (LTAO)} aims to balance short-term behaviors with long-term user interests by refining the attention mechanism. Traditional attention-based models often overemphasize recent interactions, neglecting persistent preferences. To address this, LTAO introduces a regularization term that aligns short- and long-term attention distributions, thereby enhancing content diversity over time. The regularized objective is:

\begin{equation}
    \mathcal{L}_{\text{LTAO}} = \mu \cdot \mathrm{KL} \left( \mathbf{a}_{\text{long}} \,\|\, \mathbf{a}_{\text{short}} \right),
\end{equation}

\noindent where $\mu$ is a hyperparameter controlling the regularization strength, $\mathbf{a}_{\text{short}}$ and $\mathbf{a}_{\text{long}}$ denote attention weights over recent and long-term interactions, respectively, and $\mathrm{KL}(\cdot \| \cdot)$ is the Kullback–Leibler divergence.

\textbf{Community Coverage Re-ranking (CCR)} is a post-processing strategy that promotes exposure diversity. After generating the initial recommendation list, items are re-ranked based on their marginal contribution to community coverage—favoring content from underrepresented communities. The adjusted ranking score is defined as:

\begin{equation}
    \hat{s}_i = s_i + \gamma \cdot \left(1 - \frac{n_{c(i)}}{|\mathcal{R}|} \right),
\end{equation}

\noindent where $s_i$ is the original score of item $i$, $\hat{s}_i$ is the re-ranked score, $c(i)$ is the community to which item $i$ belongs, $n_{c(i)}$ is the number of items from community $c(i)$ in the current list $\mathcal{R}$, and $\gamma$ controls the adjustment strength.

\textbf{Community Penalty Factor (CPF)} adjusts item scores to prevent dominance by overrepresented communities. During post-processing, scores are penalized proportionally to the frequency of each item's community, promoting structural diversity. The adjusted score is computed as:

\begin{equation}
    \hat{s}_i = s_i \cdot \left(1 - \alpha \cdot \frac{n_{c(i)}}{|\mathcal{R}|} \right),
\end{equation}

\noindent where $\alpha \in [0, 1]$ is a penalty weight controlling the suppression degree.

\begin{table}[ht]
\centering

\caption{Comparison of mitigation results in two datasets}
  \label{tab:mitigation strategies}
\resizebox{\linewidth}{!}{
\begin{tabular}{c | c c c c c}

\toprule
\textbf{MIND} & N@20↑ & H@20↑ & R↓& D↓ &O↑\\
\cmidrule{1-6}
Original & 5.0536 & 1.6676 & 0.9845 & 0.0015 &0.3428 \\
\cmidrule{1-6}
EGS & 5.0388 & 1.8549 & 0.9843 & 0.0015 & 0.4229 \\
Improv.\% & -0.29\% &11.23\% &0.03\% &0.14\% &23.35\% \\
\cmidrule{1-6}
CDR & 5.2372  & 1.7565  & 0.9840  & 0.0015 & 0.3603 \\
Improv.\% & 3.63\% & 5.33\% & 0.05\% & -2.04\%  & 5.08\% \\
\cmidrule{1-6}
LTAO & 5.4686 & 1.6684 & 0.9843 & 0.0014 & 0.3793 \\
Improv.\% & 8.21\% & 0.05\% & 0.02\% & 4.68\% & 10.64\% \\
\cmidrule{1-6}
CCR & 5.3526 & 2.2823 & 0.9808 & 0.0014 & 0.4371 \\
Improv.\% & 5.92\% & 36.86\% & 0.38\% & 7.82\% & 27.49\% \\
\cmidrule{1-6}
CPF & 5.0717 & 1.6807 & 0.9845 & 0.0014 & 0.3763 \\
Improv.\% & 0.36\% & 0.79\% & 0.00\% & 6.33\% & 9.77\% \\
\bottomrule
\toprule
\textbf{Adressa} & N@20↑ & H@20↑ & R↓ & D↓ & O↑ \\
\cmidrule{1-6}
Original & 4.9461 & 1.8959 & 0.9691 & 0.0020 & 0.6139 \\
\cmidrule{1-6}
EGS & 4.9237 & 1.9586 & 0.9645 & 0.0019 & 0.6694 \\
Improv.\%  & -0.45\% & 3.31\% & 0.47\% & 4.38\% & 9.04\% \\
\cmidrule{1-6}
CDR & 4.9589 & 1.9366 & 0.9641 & 0.0023 & 0.5594 \\
Improv.\% & 0.26\%  & 2.15\% & 0.52\% & -15.55\% & -8.89\% \\
\cmidrule{1-6}
LTAO & 4.9566 &  1.9694 & 0.9608 & 0.0021 & 0.4935 \\
Improv.\% &0.21\% & 3.88\% & 0.85\% & -5.26\% & -19.61\% \\
\cmidrule{1-6}
CCR & 4.8854 & 1.9442 & 0.9633 & 0.0018 & 0.6805 \\
Improv.\% & -1.23\% & 2.55\% & 0.59\%  & 7.84\% & 10.84\% \\
\cmidrule{1-6}
CPF & 4.9221 & 1.9597 & 0.9708 & 0.0019 & 0.6393 \\
Improv.\% & -0.49\% & 3.36\% & -0.18\% & 2.80\% & 4.13\% \\
\bottomrule

\end{tabular}
}
\end{table}
% 加粗

Table \ref{tab:mitigation strategies} shows the results, obtained with only a small fluctuation (2\%) in performance metrics including AUC, MRR, NDCG@5, and NDCG@10. Based on the above results, we find: (1) EGS introduces modest improvements in openness, particularly in Adressa, but shows slight decreases in content diversity metrics, suggesting that a small-scale exploration helps escape behavioral loops while maintaining model stability. (2) CDR consistently improves both content diversity and entropy across datasets, while effectively reducing the category repeat rate, indicating its strength in enhancing intra-list diversity and weakening short-term content loops. (3) LTAO consistently enhances entropy and openness, indicating its utility in reducing the recent behavioral biases. But its influence on community structure remains moderate. (4) CCR achieves the most significant improvements in group metrics, demonstrating that post-hoc re-ranking based on community coverage effectively mitigates structural information cocoons. (5) CPF delivers the largest gain in category diversity and openness on MIND, highlighting its effectiveness in promoting exposure to underrepresented groups. 
% Overall, the results confirm that different strategies excel at mitigating distinct aspects of the information cocoon effect. A combination of content-level and structure-level interventions may yield the most robust improvements in recommendation diversity and fairness.

\section{Related Work}
\subsection{News Recommendation}
Relevant studies on news recommendation systems are now abundant. Some works \cite{Zhang-2019,Wu-2021,qi-etal-2020-privacy,wu-etal-2022-two,UNBERT-ijcai2021p462} optimized personalized recommendation strategies from multiple perspectives based on the attention mechanism, such as user modeling, recall sorting, and privacy protection, distinguishing between user preferences and user-news interaction intensity. \citeauthor{Manoharan-2020} \shortcite{Manoharan-2020} combined fuzzy rules and reinforcement learning methods to recommend by data mining for social media \cite{Yu_2024}. Some researches \cite{Shi-2021,Wu-ijcai2021,Zheng-2021,Yang-2023} used graph neural networks to improve the performance of recommendation systems \cite{Fan_2019graphrec}, which provided accurate and personalized recommendations through word graph modeling, diversity optimization, and embedding. And several studies \cite{Lee-2020,Wang-2018-DKN,Wang-2018-ripplenet,Wang-2018-SHINE,Gao-2018} proposed the recommendation algorithm based on knowledge graph to provide users with more targeted and diversified news. Wu \cite{Wu-2022} proposed that news recommendation is not only a problem of sequence recommendation. Transformer could better handle the timeliness, variety and user interest dynamics. The related works \cite{DBLP:journals-2023,Promp-2024,LKPNR-2024}, on the other hand, showed the powerful potential of large language models in news recommendation, combining knowledge graph, generative models and prompts to improve the personalization and diversity of recommendations.

%\subsection{Diverse Recommendation}
%Diverse recommendation refers to improving the diversity of recommendation lists while ensuring high accuracy. The information cocoon is mainly characterized by homogeneity improvement, therefore, diversified recommendation is usually regarded as one of the most intuitive ways to mitigate the information cocoon. In an early work, Zhou \textit{et al.}\cite{pnas-2010} proposed a similarity-based algorithm to enhance the diversity of recommendations by introducing similarity weights among nodes in the user-item network, providing an effective balance between diversity and accuracy. Since then, related works have started to propose some post-processing methods based on the backbone model to improve the diversity of recommendations. For example, Romain \textit{et al.}\cite{Romain-2019} introduced TDPP with the help of the tensor decomposition technique, which combined tensor decomposition and deterministic point processes to merge diversity and accuracy into the recommendation process. Zheng \textit{et al.}\cite{Zheng-2021} proposed the recommendation system based on the graph convolutional network(DGCN), which transformed the recommendation problem as a node classification task on a graph and performed feature learning for diversity optimization. Yang \textit{et al.}\cite{DGRec-2023} used a diversity embedding generation technique to balance accuracy and diversity.

\subsection{Information Cocoon}
%Current research on information cocoons focuses primarily on detection and mitigation tasks, exploring aspects such as diversity, novelty, or the dimensions of social networks.

\subsubsection{Detection of the Information Cocoon}
Regarding the detection of information cocoons, Avin \cite{Avin2024} quantified the homogenization of social media spreading process based on networks. Michiels \cite{Michiels-2023} found a slight reduction in topic variety over time based on the data from news websites. Li \cite{Li-2022short_video} examined the information cocoon on short-form video platforms, revealing the video content and algorithmic interactions on user homogeneity. Wang \cite{WANG2024101216} defined group polarization through the use of group sentiment polarity, based on text analysis. Some Works \cite{Zhang_2024,Anwar-2024} took a comprehensive approach by considering both non-interactive diversity between users and items, as well as explicit diversity within users, defining metrics such as average type variance, historical category diversity in recommendation lists, and new category diversity. Piao \cite{Piao2023} focused on the dynamic interaction between intelligent systems and human users, exploring the emergence and development mechanisms of information cocoons from the perspective of adaptive dynamics.

\subsubsection{Mitigation of the Information Cocoon}
Existing research focused primarily on diversification control, dynamic interaction, and community polarization easing to mitigate the cocoon. 
%These studies designed and optimized the recommendation algorithms to alleviate the information cocoon effect. 
From the user perspective, one common strategy is to model diversity as a loss of regularization. Algorithms like IDSR \cite{IDSR-2020} and EDUA \cite{EDUA-2021} constructed diversity losses by capturing different user intentions, cross-category distributions of historical interactions, or varying levels of interest diversity. Works such as Zhang \cite{Zhang_2024} introduced category control parameters to propose controllable diversity frameworks. Gao \cite{CIRS-2023} proposed a counterfactual interactive recommendation system that used reinforcement learning and causal reasoning to model user satisfaction, increasing the information diversity. UCRS \cite{UCRS-2022} also used counterfactual reasoning to mitigate outdated user representations and adopted user-controllable ranking strategies to adjust recommendation diversity. From the group perspective, researchers like Antonela \cite{Antonela-2021} and Grossetti \cite{Grossetti-2019} designed community recommendation algorithms that were aware of the echo chamber effect. These algorithms aimed to recommend content and friends with differing viewpoints from the user's, promoting interactions with different communities and viewpoints to mitigate the information cocoon. Donkers \cite{Tim-2021} introduced a dual echo chamber model, incorporating cognitive and ideological factors to understand polarization in social media and explore intervention-based recommendation methods. In graph convolutional networks, the exploration of higher-order neighbors and neighborhood node aggregation, as seen in DGCN \cite{Zheng-2021}, provided ideas for mitigating group polarization by modeling interaction graph uncertainty or domain category diversity.

\section{Conclusion and Future Work}
In this study, we conducted a comprehensive assessment and analysis of the factors influencing the information cocoon effect from both individual and group perspectives. Using two real-world news datasets, we performed multiple rounds of recommendation experiments on several classic news recommendation models and assessed the associated information cocoon metrics. We examined the impact of different recommendation algorithms and designed several strategies to mitigate the information cocoon effect.

For future research, we propose to further investigate the formation and development stages of information cocoons, including the underlying mechanisms of formation, the critical emergence and transition points, and the potential mitigation conditions. We also suggest conducting further research on the dynamic trade-offs between information cocoons and task-related metrics, such as recommendation accuracy and user satisfaction. This can guide the development of more balanced news recommendation systems that better serve both user and platform needs.

\section{Acknowledgments}
This work was supported in part by the National Key Research and Development Program of China under Grant (2023YFC3310700) and  the National Natural Science Foundation of China (62572040, 62202041).

\normalsize
\bibliographystyle{ACM-Reference-Format}
\bibliography{main}

% \newpage
\appendix
\section{EXPERIMENTAL DETAILS}
\subsection{News Recommenders}\label{News Recommenders}
In this part, we present detailed information about the seven typical news recommenders used in our experiments.

\begin{table*}
\centering
\caption{Results of information cocoon metrics in multi-round recommendations.}
    \label{tab:detailed results}
\begin{tabular}{c| c c c c c c c c c }
\toprule
\textbf{MIND} & N@20 & N@50 & N@100 & H@20 & H@50 & H@100  & R & D & O\\
\cmidrule{1-10}
NAML  & \textbf{4.5611} & 10.1980 & 12.9965 & \textbf{1.2511} & \textbf{1.7258} & \textbf{2.1716} & 0.9822 & 0.0011 & \textbf{0.1502} \\
TANR  & 4.6748 & 10.2131 & 12.9965 & \underline{1.6419} & 1.9770 & 2.3709 & 0.9818 & 0.0012 & 0.2267\\
NRMS  & 5.0536 & 10.2960 & 12.9977 & 1.6676 & 1.9659 & 2.3335 & \underline{0.9845} & \textbf{0.0015} & 0.3428 \\
Hi-Fi Ark  & \underline{4.6357} & \textbf{9.1054} & 12.9986 & 1.6944 & \underline{1.9170} & \underline{2.3298} & \textbf{0.9859} & 0.0013 & 0.5863 \\
NPA  & 5.1058 & 10.2820 & 12.9973 & 1.6891 & 2.0051 & 2.3669 & 0.9831 & \underline{0.0014} & \underline{0.1821} \\
LSTUR  & 6.0568 & \underline{9.8203} & \underline{12.9900} & 2.7855 & 2.8429 & 2.9890 & 0.9600 & 0.0013 & 0.6896\\
DKN  & 7.3984 & 11.1021 & \textbf{12.9816} & 2.8459 & 2.8984 & 2.9943 & 0.9332 & 0.0005 & 0.9293 \\
\bottomrule
\end{tabular}

\begin{tabular}{c | c c c c c c c c c }
\toprule
\textbf{Adressa}& N@20 & N@50 & \:\: N@100 & H@20 & H@50 & H@100  & R & D & O \\
\cmidrule{1-10}
NAML & \textbf{4.0267} & \underline{5.5772} & \:\:  8.6569 & \textbf{0.9780} & \textbf{1.2085} & \textbf{2.4449} & \underline{0.9704} & \underline{0.0023} & \textbf{0.2253} \\
TANR  & 5.0238 & 6.5072 & \:\:  8.6558 & 1.9009 & 1.9798 & 2.6917 & 0.9690 & 0.0014 & 0.6467 \\
NRMS  & 4.9461 & 6.4579 & \:\:  \textbf{8.6550} & 1.8959 & 1.9742 & 2.6903 & 0.9691 & 0.0020 & 0.6139 \\
Hi-Fi Ark& 5.0470 & 6.5182 & \:\:  8.6552 & 1.9464 & 2.0196 & 2.7242 & 0.9488 & 0.0013 & 0.6612 \\
NPA   & 5.0398 & 6.5274 & \:\:  8.6557 & 1.8941 & 1.9747 & 2.6863 & \textbf{0.9789} & 0.0016 & 0.6417\\
LSTUR & \underline{4.2220} & \textbf{4.7246} & \:\:  8.6565 & \underline{1.2548} & \underline{1.3358} & \underline{2.5296} & 0.9699 & \textbf{0.0023} & \underline{0.5679} \\  
DKN  & 5.1290 & 6.4938 & \:\:  \underline{8.6551} & 1.9751 & 2.0443 & 2.7449 & 0.9388 & 0.0015 & 0.8202 \\
\bottomrule
\end{tabular}

\end{table*}

\begin{itemize}[left=0pt]
\item \textbf{NAML\cite{Wu2019NAML}}: A news recommendation model that utilizes different types of news information. News encoding employs a multi-view learning model, treating headlines, bodies, and categories as distinct views, with word-level and view-level attention mechanisms. User encoding is based on browsing history, using attention mechanisms to learn the user representation.

\item\textbf{TANR\cite{wu-etal-2019-TANR}}: A model that emphasizes the importance of news topic information for recommendations. Since not all news platforms provide complete topic labels, a topic-aware module is introduced to incorporate news topics as inputs to the model.

\item \textbf{NRMS\cite{wu-2019-NRMS}}: A news recommendation model that uses multi-head self-attention (MHSA) for both news encoding based on headlines and user modeling based on browsed news sequences. News and user vectors are generated through MHSA and additive attention.

\item\textbf{Hi-Fi Ark\cite{Liu-2019-HF}}: 
A knowledge-aware news recommendation model that incorporates external knowledge, such as knowledge graphs, with dynamic updates to improve the semantic understanding of news content and user preferences, ensuring precise recommendations.

\item\textbf{NPA\cite{Wu-2019-NPA}}: 
A news recommendation model that uses a personalized attention mechanism to dynamically assign attention to news items based on their relevance to individual users, enabling more accurate and tailored news recommendations.

\item\textbf{LSTUR\cite{an-etal-2019-LSTUR}}: A model that simultaneously learns long-term preferences and short-term interests. Long-term representations are derived from user ID embeddings, while short-term representations are based on recently viewed news, using a gate-controlled recurrent unit (GRU).

\item\textbf{DKN\cite{Wang-2018-DKN}}: A recommendation system with two core components: a knowledge-aware convolutional neural network that represents news by incorporating external knowledge with semantic information, and an attention mechanism that predicts the user click-through rate. %A recommendation system consists of two core components. One is a knowledge-aware convolutional neural network that represents news by incorporating external knowledge and combining semantic and knowledge information. The other is an attention mechanism that predicts the user click-through rate.
\end{itemize}

\subsection{Detailed Experimental Results}\label{Detailed Experimental Results}

We present the overall detailed experimental results of the indicators on each model in category and subcategory, which are shown in Table \ref{tab:detailed results}. This table presents the results of the individual-level and group-level indicators during the multiple rounds of recommendations on the datasets.

For the individual-level indicators, it includes the number of topic categories in the Top-K recommendation lists (denoted as N@K), the category information entropy of the Top-K recommendation lists (denoted as H@K), and the click repeat rate (denoted as R). For the group-level indicators, it includes network density (denoted as D) and community openness (denoted as O). 
In this context, \textbf{bold} indicates the most significant information cocoon effect within the corresponding indicator and its specific statistical values, while \textit{underlined} represents the second level.

\begin{figure*}[ht]
\centering
\subfigure[\textbf{Category@20}]{
\includegraphics[width=0.18\linewidth]{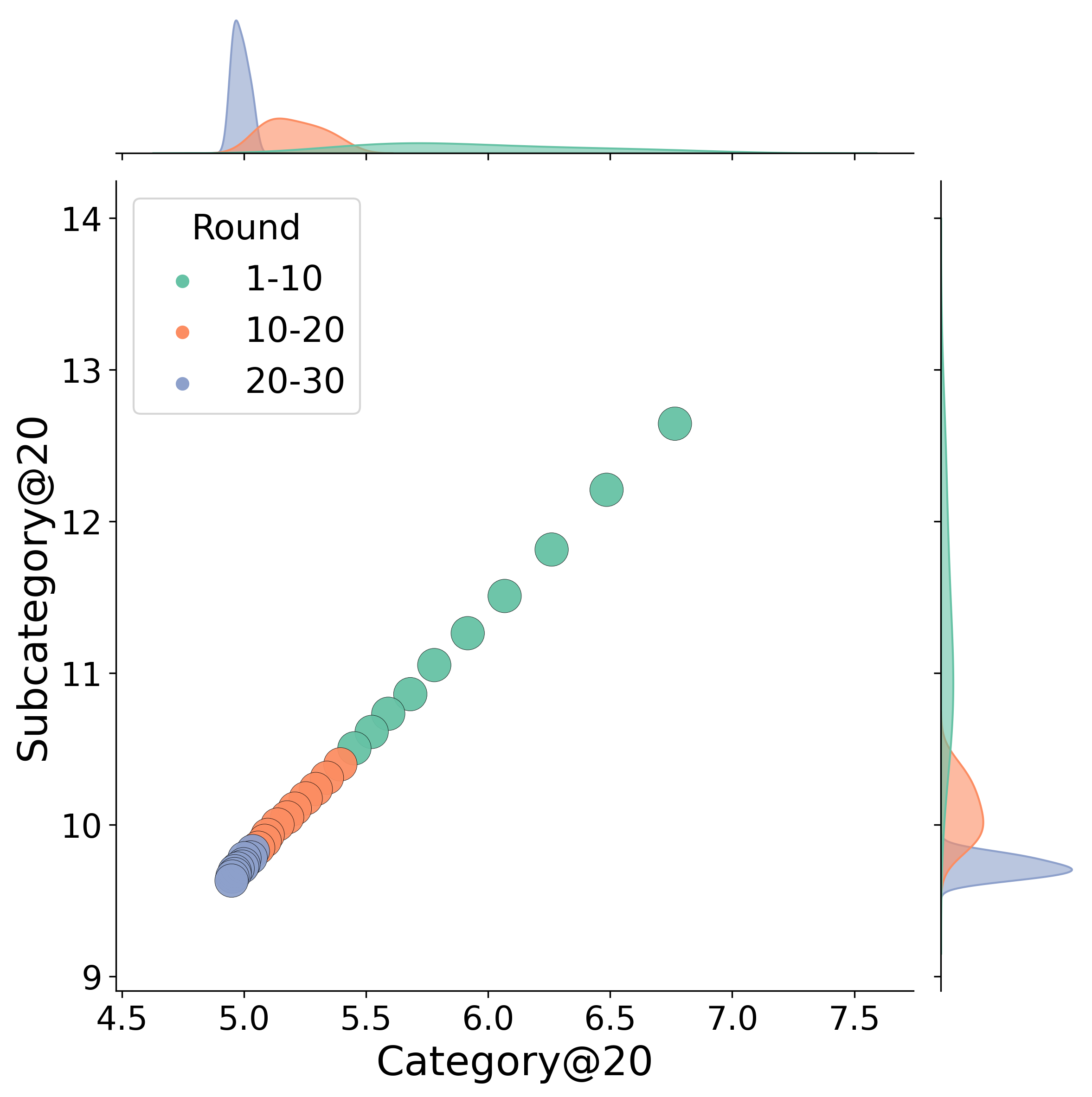}
\label{rounds_category}
}
\subfigure[\textbf{Entropy@20}]{
\includegraphics[width=0.18\linewidth]{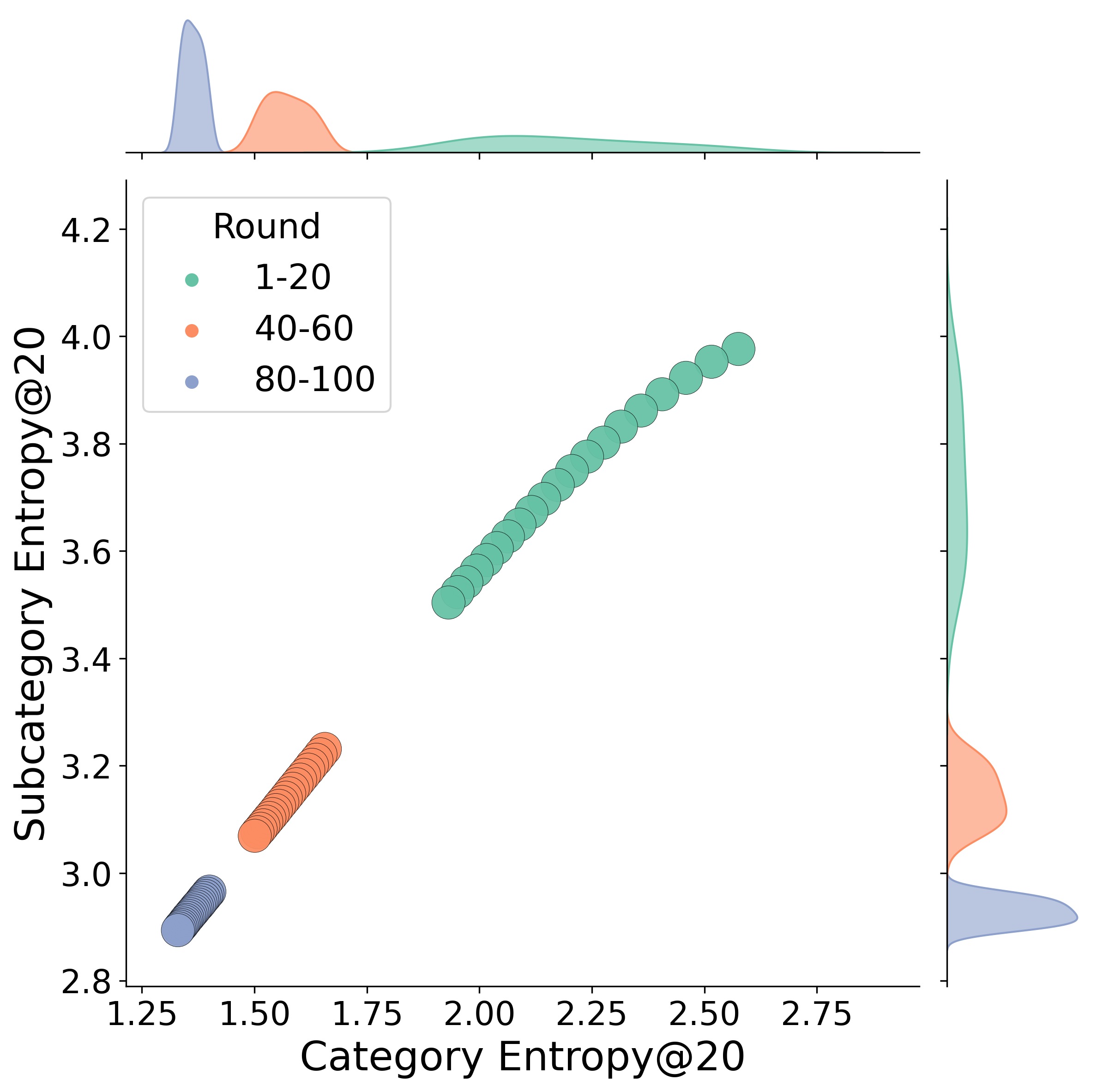}
\label{rounds_entropy}
}
\subfigure[\textbf{Click Repeat Rate}]{
\includegraphics[width=0.18\linewidth]{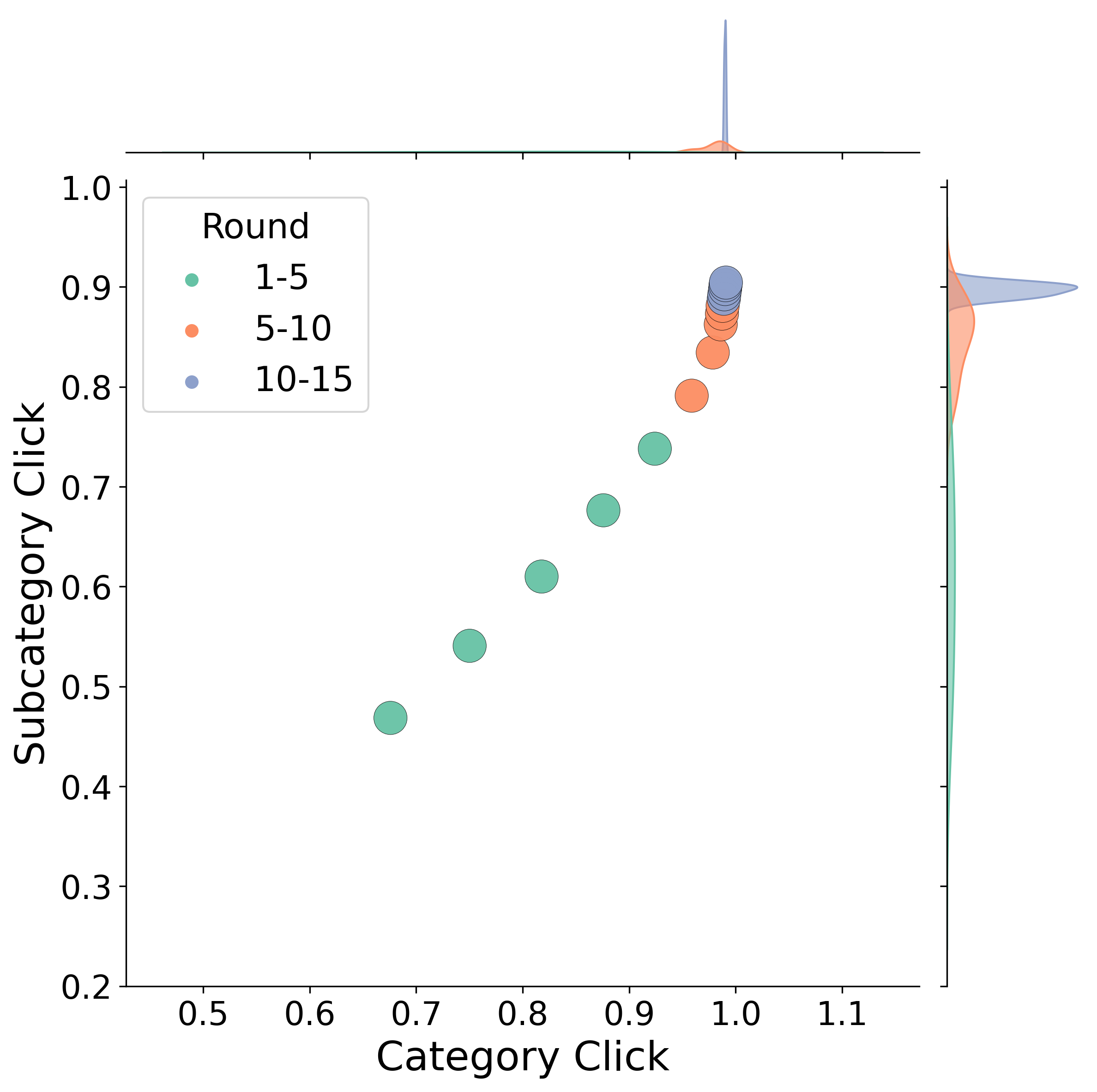}
\label{rounds_click}
}
\subfigure[\textbf{Density}]{
\includegraphics[width=0.19\linewidth]{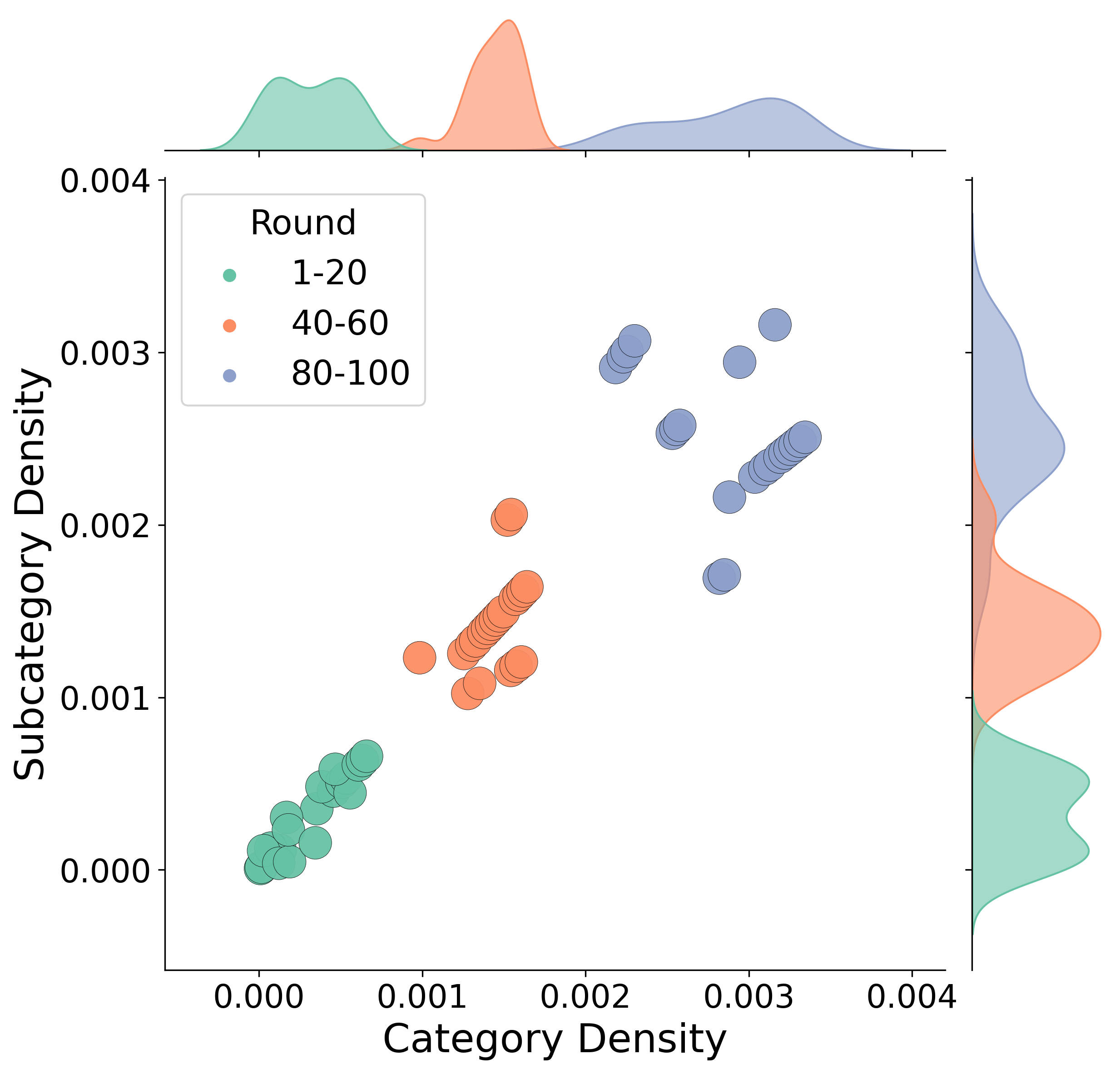}
\label{rounds_density}
}
\subfigure[\textbf{Openness}]{
\includegraphics[width=0.19\linewidth]{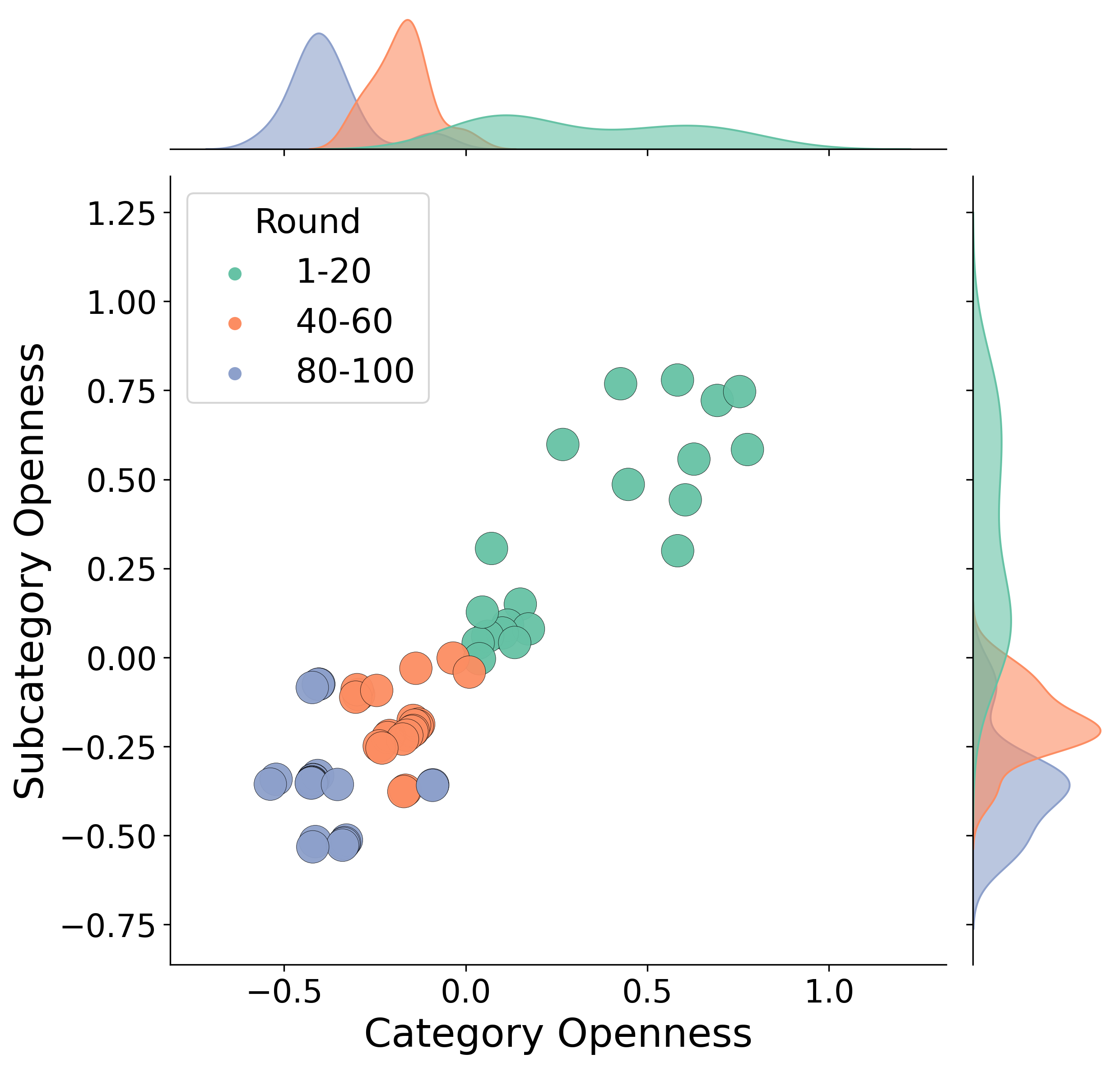}
\label{rounds_openness}
}
\caption{The histogram of the scatter distribution across different ranges of the rounds in MIND, where the x-axis shows the \textit{NRMS} model's results of the indicators under the category and the y-axis shows the subcategory.}
\label{Fig.formation}
\end{figure*}

\begin{table*}
\centering
\caption{Comparison results of categories and subcategories}
  \label{tab:detailed categories and subcategories}
\begin{tabular}{ c  |c llc llc c c }
\toprule
\textbf{MIND} & N@20&   N@50&N@100&H@20&   H@50&H@100&R& D& O\\
\cmidrule{1-10}
NAML & \textbf{4.5611} &   10.1980&12.9965 &\textbf{1.2511} &   \textbf{1.7258} &\textbf{2.1716} &0.9822 & 0.0011 & \textbf{0.1502} \\
 & \textbf{9.1109}&   23.8846 &42.6747 &\textbf{2.5144}&   \textbf{3.1429} &\textbf{3.8528} &0.8897& 0.0011& \textbf{0.2359}\\
 \cmidrule{1-10}
TANR  & 4.6748 &   10.2131 &12.9965 &\underline{1.6419} &   1.9770&2.3709 &0.9818 & 0.0012 & 0.2267 \\
 & 9.9821&   24.3634 &42.6671 &3.3832&   3.8087 &4.3530 &0.8642& 0.0011& 0.3153\\
 \cmidrule{1-10}
NRMS  & 5.0536 &   10.296 &12.9977 &1.6676 &   1.9659 &2.3335 &\underline{0.9845} & \textbf{0.0015} & 0.3428 \\
 & \underline{9.8232}&   23.8404 &42.6855 &\underline{3.2295} &   \underline{3.6549} &\underline{4.2244} &\textbf{0.9103}& \textbf{0.0014}& 0.4096\\
 \cmidrule{1-10}
Hi-Fi Ark  & \underline{4.6357} &   \textbf{9.1054} &12.9986 &1.6944 &   \underline{1.9170}&\underline{2.3298} &\textbf{0.9859} & 0.0013 & 0.5863 \\
 & 10.4033&   \underline{22.1736} &42.6849 &3.4541&   3.7793 &4.3280 &\underline{0.9021} & 0.0014& 0.6171\\
 \cmidrule{1-10}
NPA  & 5.1058 &   10.2820 &12.9986 &1.6891 &   2.0051 &2.3669 &0.9831 & \underline{0.0014} & \underline{0.1821} \\
 & 10.0241&   24.1467 &42.6757 &3.3190&   3.7498 &4.3028 &0.8876& \underline{0.0014} & \underline{0.2633}\\
 \cmidrule{1-10}
LSTUR  & 6.0568 &   \underline{9.8203} &\underline{12.9900} &2.7855 &   2.8429 &2.9890 &0.9600& 0.0013 & 0.6896 \\
 & 10.2787&   \textbf{22.1426} &\textbf{42.5962} &4.4151&   4.6389 &5.0672 &0.6382& 0.0013& 0.7413\\
 \cmidrule{1-10}
DKN  & 7.3984 &   11.1021 &\textbf{12.9816} &2.8459 &   2.8984 &2.9943 &0.9332 & 0.0005 & 0.9293 \\
 & 13.5814&   26.3178 &\underline{42.6128} &4.6780&   4.8228 &5.0734 &0.6738& 0.0055& 0.9478\\
 \bottomrule
\end{tabular}
\end{table*}

%Figure 9
\begin{figure*}[t]
\centering
\subfigure[\textbf{Subcategory@20}]{
\includegraphics[width=5.5cm,height=5cm]{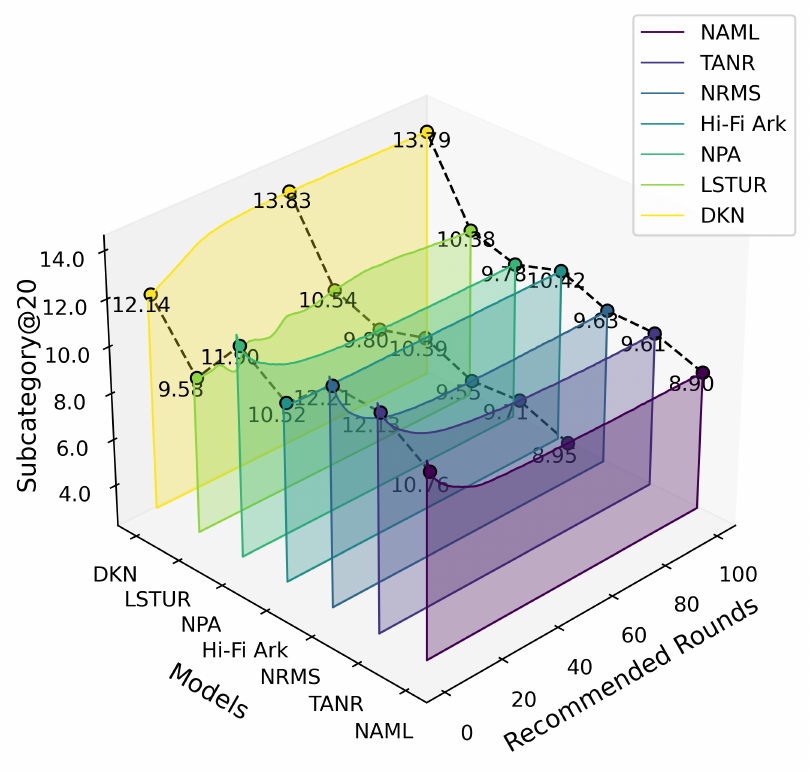}
}
\subfigure[\textbf{Entropy@20}]{
\includegraphics[width=5.5cm,height=5cm]{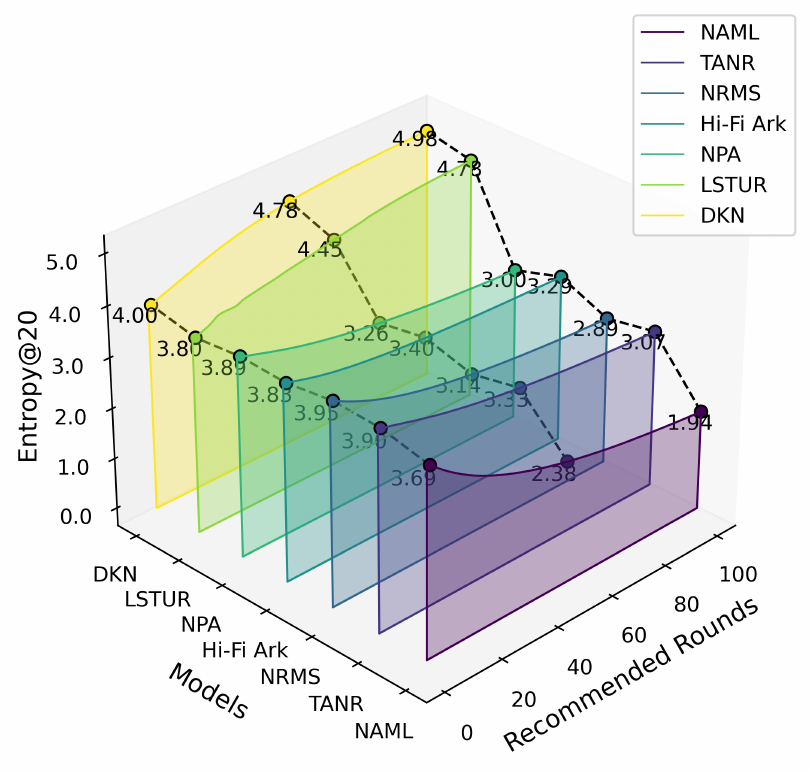}
}
\subfigure[\textbf{Click Repeat Rate}]{
\includegraphics[width=5.5cm,height=5cm]{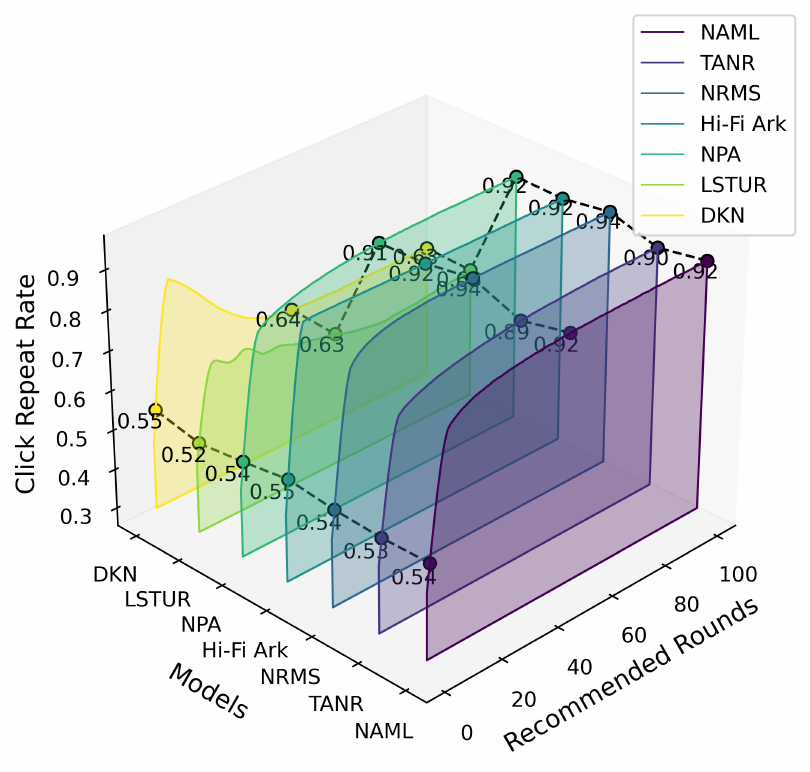}
}
\subfigure[\textbf{Network Density}]{
\includegraphics[width=6cm,height=4.5cm]{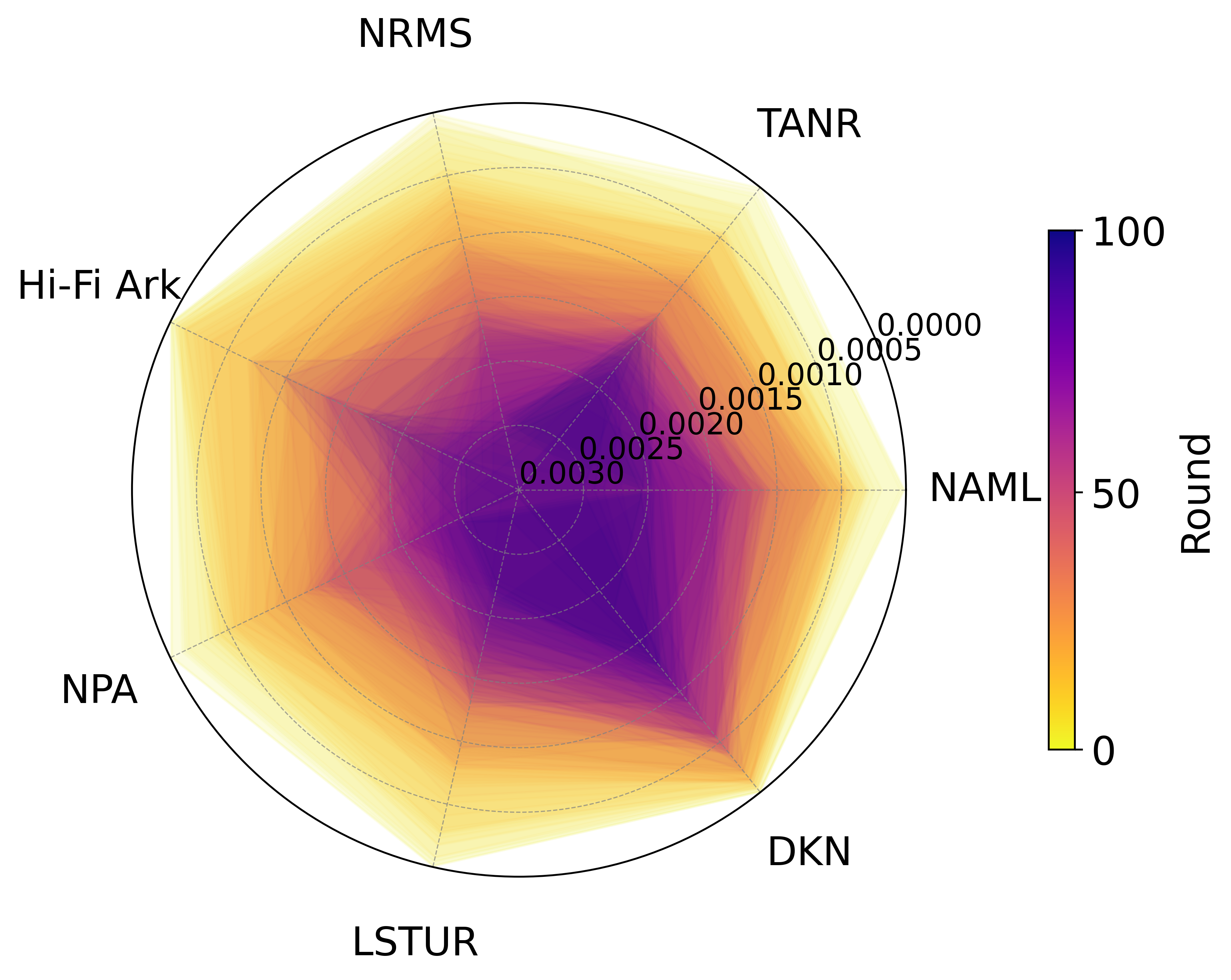}
}
\subfigure[\textbf{Community Openness}]{
\includegraphics[width=7cm,height=4.5cm]{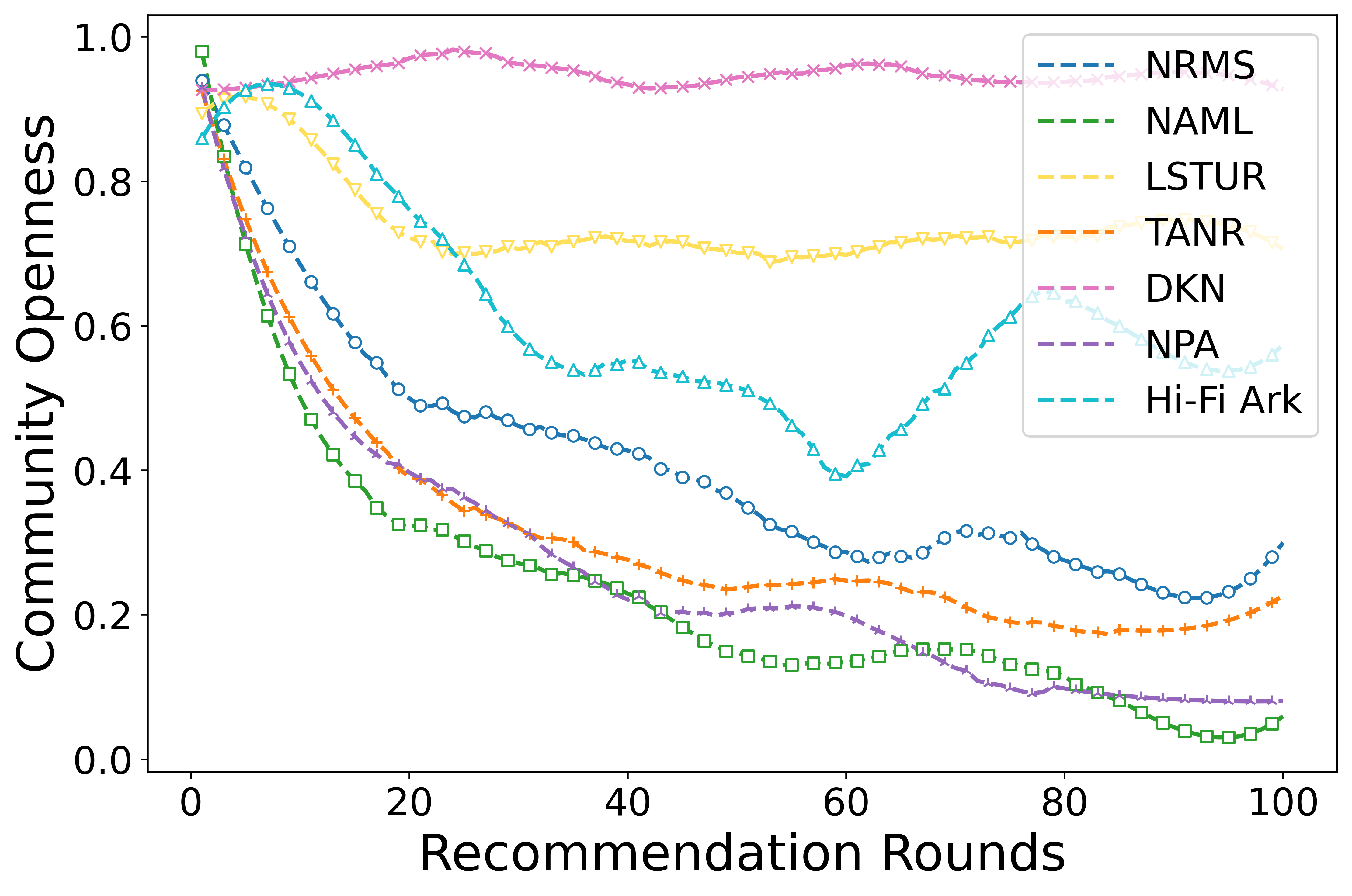}
}
\caption{The results of all metrics on subcategory in MIND after multiple recommendation rounds for each model.}
\label{Fig.subcategory}
\end{figure*}

\section{The Formation Process of Information Cocoons}

Based on the different performance of information cocoons in different recommendation rounds, we conduct further analysis on the association between the distribution of the indicator results and the recommendation rounds, and try to explore the characteristics of the formation process of the information cocoon, as shown in Figure \ref{Fig.formation}.

Positive indicators of information cocoons (click repeat rate and network density) tend to increase with the recommendation round increasing, while negative indicators (number of topic categories, category information entropy and community openness) generally decrease. These trends confirm that the information cocoon effect deepens progressively during the recommendation process.

All the scatter points show a pronounced positive correlation, indicating that the trend of the results based on the category is almost the same as that of the subcategory. This suggests that both category and subcategory reflect the deepening of the information cocoon effect as the recommendation rounds increase. Individual indicators show relatively consistent results in category and subcategory, while group indicators are more dispersed. This may indicate that, with the update of community distribution, the information cocoon effect exhibits some volatility from a group perspective.

% Furthermore, the rate of change for different indicators varies. 
For \textbf{individual-level} indicators, dramatic changes occur in the early rounds, especially the first 10, where the information cocoon effect intensifies rapidly with highly dispersed distributions. In the later rounds, the results of the indicators tend to stabilize. The number of topic categories (Figure \ref{Fig.formation}(a) becomes more concentrated between rounds 20 and 30. The category information entropy (Figure \ref{Fig.formation}(b)) reaches its stable value in the last rounds. And the click repeat rate (Figure \ref{Fig.formation}(c)) stabilizes around rounds 10-15. This difference in stabilization may be because the former two indicators are based on category, while entropy depends not only on the number of categories but also on the distribution. In the early stages, the recommendation system identifies user preferences and gradually concentrates on recommending content from specific categories, leading to a reduction in the number of categories. When the number of categories stabilizes, the distribution of top-K categories also becomes more concentrated, further reducing the diversity of recommendations. 

For \textbf{group-level} indicators, network density (Figure \ref{Fig.formation}(d)) consistently increases across different rounds. It indicates that as content becomes more targeted, the community of the user-item network becomes more concentrated, with deeper interactions within the community. Community openness (Figure \ref{Fig.formation}(e)) declines sharply in the early rounds and gradually stabilizes in later rounds. This suggests that in the early stages, recommendations primarily focus on content within the community. As the process progresses, the recommendation of external content decreases, approaching zero.

\section{The comparison of categories and subcategories}\label{The comparison of categories and subcategories}
As shown in Table \ref{tab:detailed categories and subcategories}, the information cocoon effect intensifies more rapidly for category. This can be attributed to the nature of user interests. Users tend to concentrate their behaviors within a category level, such as the topic categories of news in this study, leading to more focused interactions and a faster cocooning process. In contrast, subcategory, being finer divisions with narrower scopes, sees less frequent and less in-depth interactions, resulting in slower cocoon development. The results of all metrics on subcategory are shown in Figure \ref{Fig.subcategory}. In the group-level indicators, there is no significant difference between the results based on category and subcategory. This may be because the indicators focus on the global network. Most of the category-based indicators only show a slightly more pronounced information cocoon, which also reflects the broader and more dominant nature of user preferences within categories compared to finer subcategories.

\end{document}